\documentclass[ba, preprint]{imsart}

\RequirePackage{amsthm,amsmath,amsfonts,amssymb}
\RequirePackage[numbers]{natbib}
\RequirePackage[OT1]{fontenc}
\RequirePackage{multirow,bigdelim}
\usepackage{bold-extra, ctable, sparklines}
\usepackage{bm, graphicx, import}
\usepackage{float}
\usepackage{bbold}
\usepackage{placeins}
\usepackage{mathtools}
\usepackage{chngcntr}
\usepackage{ulem}
\counterwithin{table}{section}
\usepackage{afterpage}
\usepackage{pdflscape,array,booktabs}
\usepackage{lipsum,rotating,flafter}
\usepackage[toc,page]{appendix}
\usepackage{threeparttable,tablefootnote}
\usepackage{soul, footnote}
\usepackage[pass]{geometry}
\graphicspath{{./images/}}
\usepackage[hyphens]{url}
\RequirePackage[colorlinks,citecolor=blue,urlcolor=blue]{hyperref}
\usepackage{lscape}
\usepackage{changepage}

\newcommand{\ssnum}[1]{\textsuperscript{\citenum{#1}}}
\newcommand\Mycite[1]{\citeauthor{#1}~(\citeyear{#1})}
\newcounter{tableeqn}[table]

\makesavenoteenv{tabular}
\makesavenoteenv{table}

\counterwithin{table}{section}

\begin{document}

\begin{frontmatter}
\title{Latent Causal Socioeconomic Health Index}

\begin{aug}
\author{\fnms{Swen} \snm{Kuh}\thanksref{D,B}\ead[label=e1]{swen.kuh@adelaide.edu.au}},
\author{\fnms{Grace S.} \snm{Chiu}\thanksref{C,B,A}\ead[label=e2]{gschiu@vims.edu}}
\and
\author{\fnms{Anton~H.} \snm{Westveld}\thanksref{A}\ead[label=e3]{anton.westveld@anu.edu.au}}

\runauthor{Kuh, Chiu \& Westveld}

\address[D]{The University of Adelaide, Monash University \printead{e1}}
\address[B]{William \& Mary}
\address[C]{Virginia Institute of Marine Science; University of Waterloo; University of Washington \printead{e2}}
\address[A]{The Australian National University; Virginia Commonwealth University \printead{e3}}

\end{aug}

\begin{abstract}
This research develops a model-based LAtent Causal Socioeconomic Health (LACSH) index at the national level. Motivated by the need for a holistic national well-being index, we build upon the latent health factor index (LHFI) approach that has been used to assess the unobservable ecological/ecosystem health. LHFI integratively models the relationship between metrics, latent health, and covariates that drive the notion of health. In this paper, the LHFI structure is integrated with spatial modeling and statistical causal modeling. Our efforts are focused on developing the integrated framework to facilitate the understanding of how an observational continuous variable might have causally affected a latent trait that exhibits spatial correlation. A novel visualization technique to evaluate covariate balance is also introduced for the case of a continuous policy (treatment) variable. Our resulting LACSH framework and visualization tool are illustrated through two global case studies on national socioeconomic health (latent trait), each with various metrics and covariates pertaining to different aspects of societal health, and the treatment variable being mandatory maternity leave days and government expenditure on healthcare, respectively. We validate our model by two simulation studies. All approaches are structured in a Bayesian hierarchical framework and results are obtained by Markov chain Monte Carlo techniques.
 \end{abstract}

  
\end{frontmatter}

\vspace{-1em}

\section{Introduction}
The gross domestic product (GDP) has been conventionally used as a measure when benchmarking different countries' growth and production. However, the commonly used GDP arguably only captures one aspect/perspective---the economic performance of a country---rather than a country's \textit{overall} performance and wellbeing. Consequently, many ongoing discussions and much effort have been made to find an alternative `wellbeing' indicator as a holistic measure of a country's socioeconomic health [\Mycite{conceiccao2008}]. Such wellbeing indices are useful for governments and organizations to benchmark a country's overall performance (other than solely economic) and help policy makers form evidence-based decisions. Despite that, there are issues with existing methods that attempt to quantify this health/wellbeing feature. For instance, multiple sources of subjectivity are combined and arbitrarily turned into a single score, yet without rigorously quantifying the uncertainties around the score [\Mycite{NEF2018}; \Mycite{OECD18}; \Mycite{UN2018}], or a country's wellbeing is taken as a chosen proxy such as the life satisfaction score [\Mycite{sachs2018}], which is not a direct measurement of the variable of interest. Health and wellbeing are increasingly being accepted as multidimensional concepts that often involve multiple subjective and objective measures on the macro- and micro-levels [\Mycite{mcgillivray2006}; \Mycite{yang2018}]. We recognize that the concept of wellbeing is inevitably subjective and we focus on reducing the subjectivity on the quantifiable measures through statistical inference of the country's socioeconomic health as a model parameter.

This paper proposes a LAtent Causal Socioeconomic Health (LACSH) index by developing a hierarchical, latent variable framework to simultaneously model each country's health as a latent parameter, account for spatial correlation among countries, and evaluate the causal impact of a policy variable on the \textit{latent} health. This new methodology contributes to the aforementioned effort towards a holistic approach by addressing the subjectivity and uncertainty propagation through a single statistical inferential framework. To assess societal health for countries, the LACSH index builds on the concept of the latent health factor index (LHFI) for modeling the underlying ecosystem health in \Mycite{chiu2011} and \Mycite{chiu2013} as unobservable and latent.

Note that the approach to measuring latent traits is not unique, as the idea appears in item response theory (IRT) in the psychometrics literature [\Mycite{rabe2004}]. Other examples include the quantification of the position of political actors on a political spectrum [\Mycite{jackman2001}; \Mycite{martin2002}], constructing measures of nations' underlying democracy [\Mycite{treier2008}], and assessing ecological/ecosystem health [\Mycite{chiu2011}; \Mycite{chiu2013}]. \Mycite{rijpma2016} model the wellbeing of countries also as a latent variable, similar to the special-case LHFI model that regresses health indicators on $H$ alone. In contrast, the general LHFI model further regresses $H$ on covariates that are chosen due to their perceived explanatory nature to health. In this paper, our holistic framework further incorporates causal modeling and spatial structures into the LHFI framework. 

In applying our work, we quantify the latent health of the countries using data collected at the national level. Observable variables (e.g. gross national income (GNI) per capita, life expectancy, mean years of schooling) are treated as either \textit{indicators} or \textit{drivers/covariates} of a country's underlying health condition as opposed to \textit{measures} of health. We use `health' and `wellbeing' interchangeably to capture the notion of a country's socioeconomic performance from the social, political, economic and environmental perspectives simultaneously. For the rest of the paper we will continue to refer to this holistic notion as (latent) health when referring to \textit{both} the model parameter and the concept of wellbeing. As national-level variables tend to be spatially dependent [\Mycite{ward2018}], we incorporate a spatial modeling structure into the LHFI framework to formally model this dependency among the countries. Here, in addition to capturing latent health, we are interested in the counterfactual: what might happen to the country's health had policy makers altered the value of the policy variable in question?

To this end, we consider propensity score adjustment for reducing confounding bias in observational studies. This technique has been used widely in the literature since the seminal paper by \Mycite{rosenbaum1983}. Subsequently, there have been ample discussions [\Mycite{an2010}; \Mycite{mccandless2010}; \Mycite{kaplan2012}; \Mycite{zigler2013}] on modeling the uncertainty associated with the inference of the propensity score, as reflected by \Mycite{mccandless2009} who model the uncertainty under a Bayesian framework to evaluate the impact of statin therapy on mortality of myocardial infarction patients. We extend this idea of using the generalized propensity score framework for continuous treatment to estimate a dose-response function [\Mycite{Hirano2004}; \Mycite{imai2004}].  We evaluate the impact of varied doses of a `policy treatment' variable (for illustration purposes, mandatory maternity leave (MML) days and domestic general government health expenditure (GGHE) per capita, respectively) on a country's health. Including this notion of `policy treatment' in our model allows a model-based assessment of the effect of a policy variable on the (latent) health of a country, in the context of counterfactuals.

To elaborate on the above elements, our paper is laid out as follows. In the next section, we briefly review the methodologies used to construct some of the existing socioeconomic health indices. Section \ref{sec:data} introduces the countries' data, and Section \ref{sec:method} discusses the methodology and building blocks we employ to construct our latent socioeconomic health for nations. In Section \ref{sec:LHW}, we propose a framework (using the building blocks discussed in Section \ref{sec:method}) which is applied to the data, and highlight some of the results from our models; we also discuss a new visualization technique for assessing covariate balance under the generalized propensity score framework. In Section \ref{sec:discussion}, we revisit the data by providing an in-depth discussion of the specifics of the data and model structure we have used. In Section \ref{sec:simStudies}, we discuss two simulation studies to validate our model results. Finally, we review the limitations of our work and conclude the paper by discussing some potential future work in Section \ref{sec:future}. Appendices A--G in the supplemental document contain further details on computation, posterior distributions, and additional insights. 
\vspace{-0.7em}

\section{Review on existing indices}
\subsection{Global and regional indices}
There is an increasing awareness that the GDP has been inappropriately used as a broader benchmark measure for overall welfare among countries [\Mycite{ida2014}]. Several methods have been proposed as an alternative measure to the GDP, but existing approaches have used, e.g. the life evaluation score or `happiness' as a proxy measure of a country's health (or subjective wellbeing) [\Mycite{sachs2018}; \Mycite{conceiccao2008}]. This is also problematic, as a country's health is a multidimensional concept as aforementioned. We review five such alternative indices in Table \ref{tab:indices}. The background and components contributing to these five and other indices have been discussed by \Mycite{hashimoto2018} and \Mycite{ida2014}, but here we focus on the statistical methodology being used. Note that 2 out of these 5 indices assume equal weighting of pre-specified variables that contribute to a country's health. There appears to be little justification apart from convenience, that the concept of health should be represented by equal parts of a wide variety of variables.
\vspace{-0.5em}

\begin{table}[htb!] \label{tab:indices}
  \begin{tabular}{|p{15em}|p{20em}|}
    \hline
    \textsc{\bfseries Index} & \bfseries{\scshape Statistical Methodology} \\
    \hline
    \textbf{United Nations Human Development Index (HDI)}\ssnum{UN2018}& Arithmetic means of different variables are computed, then a geometric mean of the arithmetic means is computed to form the HDI \\
    \hline
    \textbf{World Happiness Report}\ssnum{sachs2018} & Pooled ordinary least squares regression (from econometrics) of the national average response to the survey question of life evaluations on 6 categories of variables hypothesized as underlying determinants of the nation's `happiness score'\\
    \hline
    \textbf{Social Progress Index (SPI)}\ssnum{stern2020} & First, a principal component analysis (PCA) is used to determine the weighting of indicators within each component, and the weights and indicators are multiplied to obtain component scores. Next, component scores are transformed onto a scale of 0-100, an arithmetic mean is computed for each dimension, and another arithmetic mean is computed to obtain the final SPI \\
    \hline
    \textbf{Happy Planet Index (HPI)}\ssnum{NEF2018} & The variables `experienced wellbeing' and `life expectancy' are multiplied, then divided by `ecological footprint'; scaling constants are used to map the final HPI to range from 0-100  \\
    \hline
    \textbf{Organisation for Economic Co-operation and Development (OECD) Better Life Index (BLI)}\ssnum{OECD18} & OECD BLI website user-specified weights are assigned to each topic (e.g. education, income), and up to four indicators which constitute each topic are assigned equal weights to form the final BLI \\
    \hline
  \end{tabular}
  \caption{Selection of existing indices and methodology used}
\end{table}
\vspace{-1.5em}

\section{Data}\label{sec:data}
The variables in this work are chosen based on data availability and quality, and in alignment with other existing well-being indices. They can be roughly grouped according to environmental and infrastructure, progressiveness and inclusiveness, education, income and wealth, and demographic variables. We collated our data from the years 2010--2016 from publicly available databases. We consider 15 metrics, $\bm{Y}$, each of 2 treatment variables, $T$, and 4 covariates, $\bm{X}$, shown in Table \ref{tab:variables}. 

Most of the metrics and covariates employed in our models are taken from the data section in the United Nations Human Development Report, which is sourced from various organizations and the World Bank database. Specifically, the POLITY variable is sourced from the Polity IV project [\Mycite{polity}], and Corruption Perception Index from the Transparency International website [\Mycite{CPI}]. Other relevant variables (e.g. literacy rate among adults in the country) were not included in our model due to a substantial amount of missing data. 

\begin{table}[htb!]
\begin{tabular}{llc}
  \toprule
  \textbf{X, Covariates}& T, \textbf{Treatment variable} \\
  \midrule
  Forest area$^\sharp$ & Federally mandated maternity \\
  Access to electricity, rural & \ \ leave (MML) days \\
  Mean years of schooling & Domestic general government expenditure \\
  Population, total$^\flat$ & \ \ on health (GGHE) per capita$^\flat$ \\
  \midrule
  \multicolumn{2}{c}{	\textbf{Y, Metrics}}  \ML
  Education index & $^\star$Population density$^\flat$ \\
  $^\star$Popn., urban (\% of total) & $^\star$Popn., ages 65 and older$^\flat$ \\
  $^\star$Employment to popn. ratio (\%) & $^\star$Unemployment rate (\%)$^\sharp$ \\
  $^\star$Corruption Perception Index$^\sharp$ & Life expectancy$^\partial$ \\
  Infant mortality rate$^\sharp$ & Internet users (\% of popn.)$^\sharp$ \\
  $^\star$Renewable energy consumption (\%)$^\sharp$\footnote[2]{At the time that this manuscript was being prepared, the latest publicly available data obtained on this metric was from 2015.} & $^\star$POLITY index$^\partial$ \\
  Gross National Income (GNI)\NN \ \ per capita (current international \$)$^\flat$  &  \\
  $^\star$Prop. of parliamentary seats \NN \ \ held by women (\%)$^\sharp$ & \\ 
  Popn. with at least some \NN \ \ secondary education (\% ages 25 and older) & \\
  \bottomrule
  \multicolumn{2}{r}{ 
  \textit{\scriptsize $^\star$Variables included as $Y^*$} \
  \textit{\scriptsize $^\sharp$square-root transformed} \
  \textit{\scriptsize $^\flat$log-transformed}} \
  \textit{\scriptsize $^\partial$cubic-transformed}
\end{tabular}
\caption{List of variables $\bm{X}, {T} \text{ and } \bm{Y}$}
\vspace{-0.5em}
\label{tab:variables}
\end{table}

Literature on Neyman-Rubin's causal framework advocates the use of pre-treatment variables to answer causal questions. For our case studies, we argue that unlike the policy variables, existing traits on infrastructure and demography are what a country ``comes with''. These ``inherent'' traits are therefore regarded as \textit{drivers} of socioeconomic health, and as pre-policy-treatment covariates. Specifically, the $Y$'s are \textit{indicators} of health (e.g. education index) based on measures that we perceive as \textit{reflective} of a country's health and regarded as the outcome variables due to various policies. In particular, GNI as opposed to GDP was used as it is perceived as a more inclusive indicator of a country's wealth [\Mycite{klugman2011}]. These indicators, or metrics, have been \textit{a priori} transformed so that increasing values reflect better health and to reduce skewness; see Section \ref{ssec:missing} for additional details. For our single policy treatment variable $T$, we consider each of the following in two separate models --- federally mandated number of maternity leave (MML) days and domestic general government health expenditure (GGHE) per capita. These two variables were chosen based largely on data availability, but also on the fact that their proposed economic and societal benefits [\Mycite{chapman2008}; \Mycite{lea1993}] make them interesting examples to illustrate our methodology. Note that the World Bank data source only has alternate years of maternity leave data, and we had to informally impute the data for some of the OECD countries using data from the OECD website [\Mycite{OECD}]. A discussion of data imputation is found in Section \ref{ssec:missing}. 

Importantly, it is recognized that the selection of modeled variables is inevitably subjective but still in line with other literature and well-being indices. As such, we focus on the methodology and its interpretation.

\section{Methodology}\label{sec:method}
To quantify the \textit{latent causal socioeconomic health} (LACSH) and its uncertainty in a policy-specific context, we integrate two modern approaches---the latent health factor index (LHFI) [\Mycite{chiu2011}] and the generalized propensity score (GPS) methodology and its extensions [\Mycite{Hirano2004}, \Mycite{imai2004}]---along with spatial modeling to account for spatial dependence among countries. We are interested in these two methods as the former describes health as \textit{latent}, i.e. a trait that is not directly measurable, while the latter allows us to examine the effect of policy prescription and estimate the dose-response function for different `doses' of a policy treatment and the corresponding response---a nation's overall socioeconomic health---as a counterfactual.

\subsection{Latent health}\label{ssec:LHFI}
\begin{figure}[H]
\begin{center}
  \includegraphics[width=0.7\textwidth]{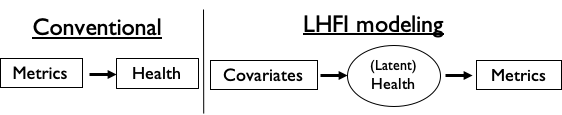} \vspace{-1.2em}
\end{center}
\caption{Schematics of a conventional health index and LHFI [adapted from \Mycite{chiu2013}]}
\label{fig:LHFI}
\end{figure}

As an analogy to a country's latent health, the underlying health conditions of a person who is deemed healthy cannot be directly compared to those conditions of another person. It is the measurable variables such as height, weight or calorie intake of a person that can be compared. Similarly, for a country, there is no single directly observable quantity that can represent ``how well a country is doing". Thus, the health of a country is a notion that we wish to evaluate comprehensively and holistically. For instance, we may argue that variables like GNI, life expectancy, and infant mortality rate can each coarsely inform us on some aspect of the state in which a country's health is, but not its \textit{overall} health. The LHFI framework unifies multiple aspects of health by modeling the underlying condition that we wish to assess as a \textit{latent} parameter (not directly measurable), but it is dependent on different measurables that are either drivers of health (covariates), or indicators of health (metrics); see Figure \ref{fig:LHFI}.

\vspace{-0.5em}
\vspace{-1.5em}

The LHFI structure employed to model health as a latent variable for our specific context is a type of mixed model [\Mycite{rabe2004}], where nation-specific health is a  random effect. We formulate our model as a Bayesian hierarchical mixed model, as it is noted in \Mycite{gelman2007} as the most direct approach to handle latent structures. 

\Mycite{martin2002} discuss the unidentifiability issues that are prominent in item response models (a type of generalized linear mixed model) in their work. To address this issue in our framework, we truncate the distribution for one pre-selected country's health parameter, $H_{anc}$, to anchor our latent score's scale. A more in-depth discussion of the anchoring approach can be found in Section \ref{ssec:ident}. 

Note that a country's metrics are multivariate in nature. Thus, in our hierarchical model, we use a multivariate normal distribution on the first level in the hierarchy, i.e. the metric level (Y-level) (equation (\ref{eq:MVNy})). We consider the previous years' covariates and metrics, averaged over years, as a set of combined covariates that drive the countries' health in the following year, which, in turn, is reflected by that year's metrics. 

Specifically, we designate the year 2015 as the current year, in which the treatment is administered, and we evaluate its effect on the country's socioeconomic health and metrics in the following year (2016).  Pre-treatment covariates are $\bm{X}^*$ and $\bm{Y}^*$, where $\bm{X}^*$ denotes the averaged covariate values over the years 2010--2014. Similarly, $\bm{Y}^*$ denotes the averaged metric values over the years 2010--2014; to avoid collinearity among regressors, we retain only one of the $Y^*$'s that show a correlation of 0.8+ with another $Y^*$ or an $X^*$ (see Table \ref{tab:variables}). In particular, we eliminate the metrics one-by-one until all correlations between $\bm{X}^*$ and $\bm{Y^*}$ are less than 0.8. Both $\bm{X}^*$ and $\bm{Y}^*$ are regarded as predictors of latent health in 2016. (The exclusion of 2015 from the definitions of $\bm{X}^*$ and $\bm{Y}^*$ will be further discussed in Section \ref{ssec:BPSA}.) Therefore, our base LHFI model (excluding spatial elements) with an `$H$-anchor' takes on the form 
\vspace{-0.2em}
\begin{eqnarray}
  \bm{y_i} | \bm{a}, H_i, \Sigma_Y & \stackrel{ind.}{\sim}& \text{MVN}(\bm{a}H_i, \Sigma_{Y}) \label{eq:MVNy} \\
  \bm{H} | \bm{\zeta}, \bm{W^*}, \sigma^2_H & \sim & \text{TMVN}( \bm{W^*}\bm{\zeta}, \Sigma_H) \mathbb{1}\{H_{anc} < 0 \} \label{eq:MVNh}        \\
  \text{where}  \ \ \
  \Sigma_H                                            & = & \sigma^2_H{\boldsymbol {I}} \notag 
\end{eqnarray} \vspace{-1.3em}

\noindent and where MVN and TMVN denote the multivariate and truncated multivariate normal distributions, respectively. The TMVN is defined as the joint distribution of $N-1$ MVNs (for non-anchor countries) with a truncated normal (for $H_{anc}$). This joint distribution has mean $\bm{W^* \zeta}$ and an $N \times N$ diagonal covariance matrix $\Sigma_H$, which reflects the naive assumption that countries are independent given the covariates. See \Mycite{horrace2005} for full details of the TMVN formulation. Normality is assumed due to the nature of our metric variables (see Section \ref{sec:data}).
 
At the Y-level, we let $\bm{y_i} = (y_{i1}, \ldots, y_{iP})^T$ be a $P \times \text{1}$ vector for the $i$th country's metrics in the year 2016 for $i = 1, \ldots, N$; $\bm{a} = (a_1, \ldots, a_P)^T$ be the $ P \times 1$ vector for the `loadings' of any country's health on its metrics; and $\Sigma_Y$ be the $P \times P$ covariance matrix for the 2016 metrics. \

We refer to equation (\ref{eq:MVNh}) as the health level ($H$-level), where $\bm{H}$ is an $N \times \text{1}$ vector of \textit{latent} health in 2016 for all $N$ countries \textit{including} the chosen anchor country; $\bm{W^*} = (\bm{1}, \bm{T}, \bm{X^*}, \bm{Y^*})$ is an $N \times (2 + K + Q)$ matrix where $\bm{1}$ is an $N \times 1$ vector of ones; $\bm{T}$ is an $N \times 1$ vector of treatment values in 2015; $\bm{X^*} = (\bm{x^*_1}, \ldots, \bm{x^*_K})$ is an $N \times K$ matrix for $K$ different covariates, $\bm{x^*_k}$ being an $N \times 1$ vector of the $k$th covariate for $k = 1, \ldots, K$, averaged over the years 2010--2014; and $\bm{Y^*} = (\bm{y^*_{1}}, \ldots, \bm{y^*_{Q}})$ is an $N \times Q$ matrix where $\bm{y^*_q}$ is an $N \times 1$ vector of the $q$th metric for $q = 1,\ldots, Q$, also averaged over the years 2010--2014 (and $Q \textless P$); $\bm{\zeta} = (\zeta_0, \zeta_1, \ldots, \zeta_{K+Q+1})^T$ is a $(2+K+Q) \times 1$ vector including the intercept and slope coefficients corresponding to the treatment variable, the five-year averaged pre-treatment covariates and the five-year averaged pre-treatment metrics; $\sigma^2_H$ is the common variance for $H_i$ for all $i$ including $H_{anc}$; $\boldsymbol{I}$ is an $N \times N$ identity matrix; finally, $H_{anc}$ for the chosen anchor country is restricted in $(-\infty, 0)$ (see Section \ref{ssec:ident} for justification).

Note that the raw metrics are on vastly different scales, so we standardize each metric to have mean zero and unit variance. As such, the overall intercept of the model appears at the $H$-level rather than the $Y$-level. A discussion of data transformations is found in Section \ref{ssec:missing}.
\vspace{-0.7em}
\subsection{Spatial modeling}\label{ssec:spatial}
\Mycite{ward2018} and \Mycite{darmofal2015} note that there are very few examples of point-level data in the social sciences. In addition, spatial correlations among countries are often not accounted for, even though macro-level variables of countries are expected to be spatially correlated, as countries that are close together in regions (e.g. Europe, North America and Central Asia) tend to be more similar in terms of a cultural, economic, social or political context. This suggests that the \textit{latent health} of countries may also be spatially dependent. In order to assess the need for spatial modeling in our framework, we fit the base LHFI model, treating the policy variable as a typical covariate, in this case MML, using equations (\ref{eq:MVNy}) - (\ref{eq:MVNh}) and examined its residuals. The residuals are defined as 
\vspace{-0.3em}
$$\bm{\widehat{\epsilon_H}}^* = \bm{H} - \bm{X\zeta} $$
where $\bm{\widehat{\epsilon_H}}^*$ is the posterior median of the residuals after subtraction on the right-hand-side based on the Markov chain Monte Carlo (MCMC) samples. 

\begin{figure}[h!]
  \includegraphics[width=.9\textwidth]{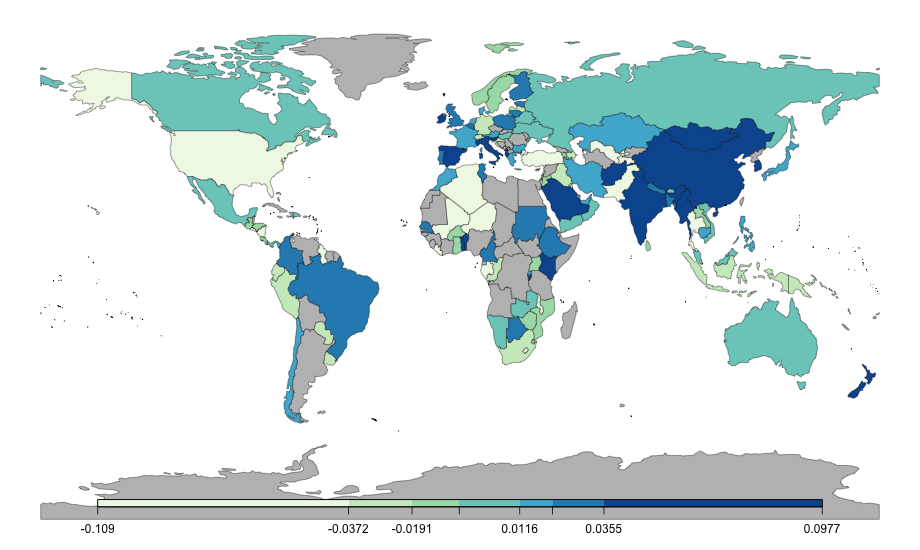} \vspace{-2em}
    \caption{Residuals $\bm{\widehat{\epsilon_H}}^*$ on the world map. Gray regions are not represented in our dataset}
    \label{fig:Hres} \vspace{-1.5em}
\end{figure} \nocite{south11} 

Figure \ref{fig:Hres} presents the residuals from the base LHFI model fit on the world map. It is apparent that countries that are geographically close together in regions such as North America, Western Europe, Central and South East Asia have residuals that are either similarly under- or over-estimated by our base model. Similarly, the model that replaces MML with GGHE (map not shown) also shows the need to address spatial dependency. To accommodate this, in the next section we incorporate spatial dependency among residuals on the health level, and modify equation (\ref{eq:MVNh}) to account for this spatial dependence in its residuals.

\subsection{Causal inference}\label{ssec:BPSA}
In addition to quantifying our latent socioeconomic health and its uncertainty, we seek to integrate causal modeling into our framework to provide insight into the effect of a `policy treatment' variable on the health of a country.

Two schools of thought dominate the causal inference literature --- namely, ``Pearl's causal diagram'' [\Mycite{pearl2009}] and ``Rubin's causal model'' [\Mycite{imbens2015}]. Both attempt to establish causal effects from observational studies, which was previously considered impossible because such studies are not randomized controlled trials [\Mycite{imbens2015}; \Mycite{hernan2018}]. Among causal inference methods for non-experimental data, propensity score (PS) analysis (stratification, matching and covariate adjustment) in the so-called Rubin's approach has been widely used to address selection bias. In our current work under the Rubin framework, instead of dichotomizing the continuous treatment variables, we consider the generalized propensity score (GPS), which were developed similarly by \Mycite{Hirano2004} and \Mycite{imai2004}, to estimate the dose-response function. The GPS approach is an extension to the propensity score method for binary treatments and multi-valued treatments [\Mycite{rosenbaum1983}; \Mycite{imbens2000}]. It allows us to fully utilize the raw information while reducing the ambiguity due to an arbitrary quantile used to categorize treatments. We follow the specifications as laid out by \Mycite{Hirano2004} in our work.

In the case of binary treatments, there has since been research that considers the uncertainty in the propensity scores [\Mycite{mccandless2009}; \Mycite{an2010}], although incorporating the outcome variable at the stage where the inference of the PS is conducted may be contentious [\Mycite{kaplan2012}; \Mycite{zigler2013}; \Mycite{zigler2016}]. For this reason, our GPS framework extends the work by \Mycite{zigler2013} for binary treatment, whereby we use the Bayesian posterior predictive distribution of the GPS to separate the design stage and analysis stage in order to `cut the feedback' (i.e. to ensure that the inference of the GPS does not depend on the outcome variable) [\Mycite{mccandless2010}; \Mycite{zigler2013}; \Mycite{zigler2014}]. Cutting the feedback is further explored by \Mycite{stephens2023}; like various discussants of this work, we take the viewpoint that cutting the feedback is fully consistent with the Bayesian paradigm. See Appendix A for implementation details. 

Irrespective of a categorical or continuous treatment variable, there are three main assumptions in Rubin's approach of causal modeling, namely, (i) the stable unit treatment value assumption (SUTVA), which stipulates no interference between units [\Mycite{rubin2005}]; (ii) strongly ignorable treatment assignment, which stipulates no unmeasured confounders [\Mycite{rosenbaum1983}]; and (iii) consistency, where the potential outcome of the treatment must correspond to the observed response when the treatment variable is set to the observed `exposure' level [\Mycite{cole2009}]. For the GPS method, \Mycite{Hirano2004} generalize (ii) to the weak unconfoundedness assumption,  which only requires conditional independence for each value of the treatment ($H(t) \perp T \ | \ X, \text{ for all } t \in T$) as opposed to joint independence for all potential outcomes.

However, incorporating causal modeling in a spatial setting potentially violates the no interference assumption (SUTVA) as discussed at length in \Mycite{keele2015} and \Mycite{noreen18}. In investigating the effect of convenience voting and voter turnout, \Mycite{keele2015} are concerned about interference and spillover effects -- the units (individuals) may be influenced due to proximity of geographical regions (or influenced at the workplace or by their social network, etc.).

In our paper, the causal question of interest is the effect of a national policy variable on a country's health, with the unit of interest being at the national-level rather than at the individual-level. We consider two policy variables in our work. For the mandatory maternity leave (MML) variable, two obvious scenarios of interference and spillover in our case may be a) individuals immigrating or emigrating and in their newly adopted country, either influencing policy makers or affecting the \textit{health} of the country (e.g. a Canadian mother whose wellbeing benefited from the Canadian federal maternity leave policy emigrates to the United States, which does not have federal maternity leave, thus possibly improving the \textit{health} of the United States); b) policy makers being influenced by their international social networks. 

The second policy variable of interest is domestic general government health expenditure (GGHE) per capita in a country. Note that `health' in this variable refers to the individual-level's public healthcare funding rather than the countries' overall socioeconomic health $H$. We argue that the potential scenarios of interference and spillover are similar to a) and b) mentioned above.

Here, we can assume minimal effects of individuals' international migration on MML, GGHE, or socioeconomic health at the national level. Additionally, we can assume that federal policy making regarding maternity leave and public healthcare expenditure is a collective domestic effort and generally conducted with minimal foreign interference. Finally, as discussed by \Mycite{schutte2014}, when there is only minor overlap in the units, a consistent treatment effect can still be valid. These arguments suggest that SUTVA is reasonable in our case.

Moreover, we utilize structural variables, namely, the country's existing infrastructure and the average of previous years' metrics as covariates in our framework. This is to align with the approach by \Mycite{rubin2005} of conditioning on the \textit{pre}-treatment variables. We can assume that the current year's policy treatment is affected by metrics and covariates from previous years. We assume that, given these observable pre-treatment covariates through the GPS, a country's choice of MML days and GGHE is random, and that there are no unmeasured important confounders. Hence, we proceed with the GPS framework while assuming that the required assumptions (i), (ii), and (iii) hold. 

\begin{figure}[!htbp]
\centering
 \includegraphics[width=0.7\textwidth]{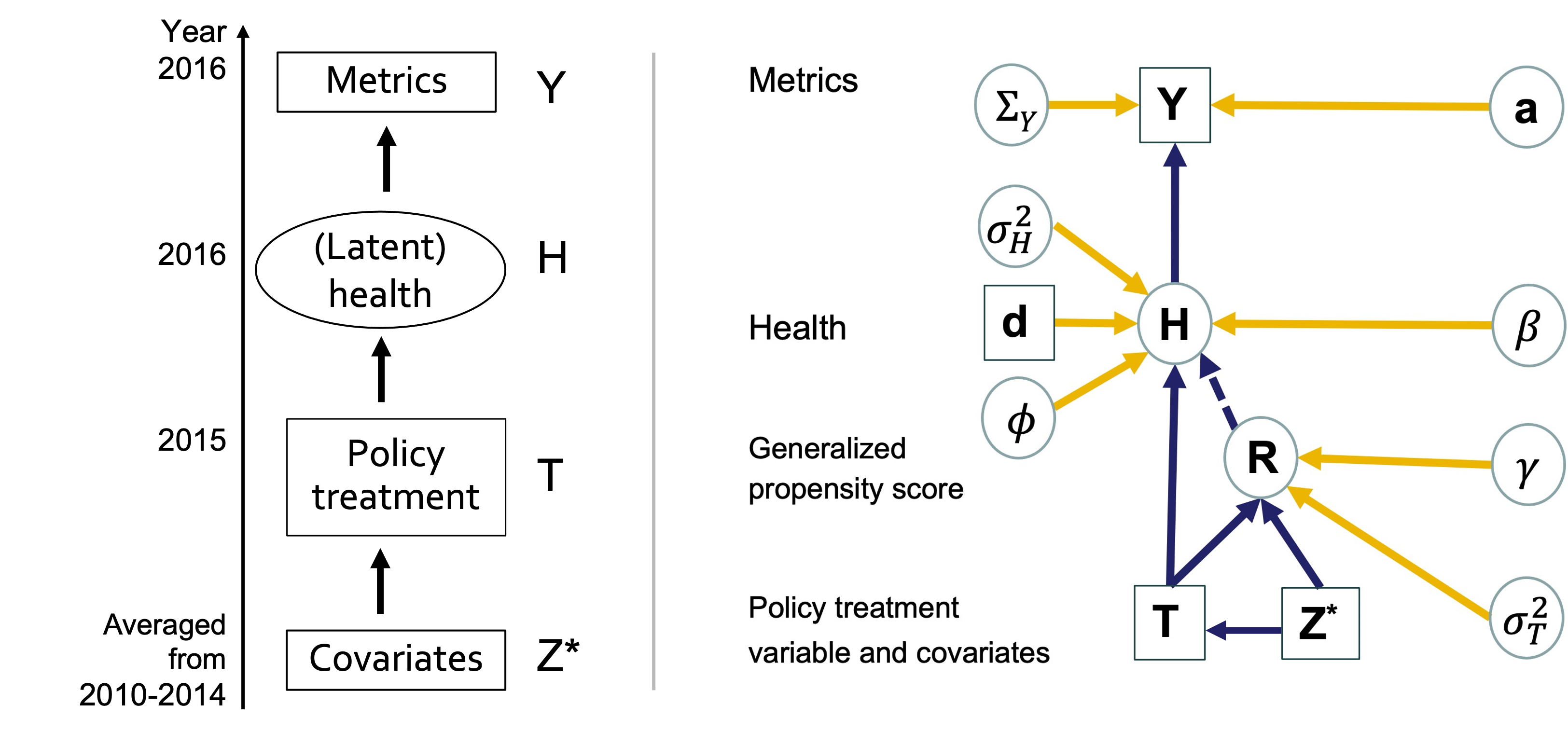} \vspace{-0.5em}
		\caption{Left panel: Extension of the LHFI framework with causal modeling. Right panel: Causal directed acyclic graph (DAG) of our full LACSH model (dark purple arrows only) and the model's inference graph (arrows in either color), where the dashed arrow indicates `cutting the feedback'.}
  \label{fig:LHFInew}
\end{figure}

Our GPS formulation is based on work by \Mycite{Hirano2004}. Similar to the PS approach for the binary case, in which the PS is the probability of receiving treatment given the covariates, the GPS is defined as the conditional probability density of the continuous treatment given the covariates. The relevant properties and methodology are discussed at length in \Mycite{Hirano2004} and \Mycite{kluve2012}. The `outcome variable' in our work is the country's latent health. A schematic representation is presented in the left panel of Figure \ref{fig:LHFInew} and we represent the full inference and causal DAG in the right panel. 

As such, we propose the LHFI methodology that includes causal and spatial modeling as the LAtent Causal Socioeconomic Health (LACSH) index. To incorporate causal modeling into our spatial LHFI model, we introduce the policy treatment variable $T$ and its generalized propensity score $R = r(T, \bm{\gamma}, \bm{Z^*}, \sigma^2_T)$ to our health-level through its mean (however, note the special MCMC implementation in Appendix A regarding ``cutting the feedback'' in the MCMC):

\vspace{-1em}
\begin{eqnarray}
\bm{y_i} | \bm{a}, H_i, \Sigma_Y &\stackrel{ind.}\sim& \text{MVN}(\bm{a}H_i, \Sigma_{Y}) \label{eq:B1} \\ 
  \bm{H} |  \bm{\beta}, \bm{T}, \bm{R}, \Sigma_H &\sim& \text{TMVN}(\bm{\mu}, \Sigma_H)\mathbb{1}\{H_{anc} < 0\} \label{eq:eq2}                 \\ \vspace{2em}
  T_{i} | \bm{Z^*}_{i}, \bm{\gamma}, \sigma^2_T &\stackrel{ind.}\sim& \text{N}(\bm{Z^*}_{i}\bm{\gamma}, \sigma^2_T) \label{eq:trt} \\
  \text{where}  \hspace{3em}
  [\bm{\mu}]_{i} &=&  \beta_0 + \beta_1 T_{i} + \beta_2 T_{i}^2 + \beta_3 R_i + \beta_4 R_i^2 + \beta_5 T_i R_i \label{eq:H_mean} \\
  R_i = r(T_{i}, \bm{\gamma}, \bm{Z^*_i}, \sigma^2_T) &=& \frac{1}{\sqrt{2\pi} \sigma_T} \text{exp} \left( - \frac{1}{2\sigma_T^2}(T_{i} -  \bm{Z^*}_{i}\bm{\gamma})^2 \right)  \label{eq:trt2} \\
  \Sigma_H &=& \sigma_H^2 \mathbf{\Omega}(d,\phi) \label{eq:3.32} \\
  \mathbf{\Omega}(d,\phi)                                  &=& \begin{bmatrix}
    1         & \rho_{12} & \cdots    & \rho_{1n} \\
    \rho_{21} & 1         & \ddots    & \vdots    \\
    \vdots    & \ddots    & \ddots    & \rho_{n-1,n} \\
    \rho_{n1} & \cdots    & \rho_{n,n-1} & 1         \\
  \end{bmatrix} \label{eq:spa4}                      \\
  \rho_{nm} &=& exp(-d_{nm}/\phi) = \rho_{mn} 
  \label{eq:spa5}
\end{eqnarray} \vspace{-2em}

\noindent where $\Sigma_H$ denotes the $N \times N$ spatial covariance matrix for \textit{health}; $\rho_{nm}$ is the correlation parameter between countries $n$ and $m$, which is a function of both $d_{nm}$ (the great circle distance (GCD) between capital cities of two countries\footnote{See \url{http://ksgleditsch.com/data-5.html}} and $\phi$ (the `range' or inverse rate of decay parameter). 

In equation (\ref{eq:H_mean}), $T_{i}$ is the policy treatment variable of interest in 2015, $R_i = r(\cdot)$ is the GPS, and $\bm{\beta} = (\beta_0, \beta_1, \ldots, \beta_5)^T$ are the associated regression coefficients. The inclusion of quadratic and interaction terms of the GPS and treatment variable are described in Section \ref{ssec:coefB}. In equation (\ref{eq:trt2}), $\bm{Z^*}_i$ is the $i$th row vector of the $N \times (K+Q)$ matrix $\bm{Z^*} = (\mathbf{1}, \bm{X^*}, \bm{Y^*})$, whose columns are the $N \times 1$ vector of ones, and the covariates and metrics averaged over 2010--2014 (as described in Section \ref{ssec:LHFI}); the corresponding $(1+K+Q) \times 1$ regression coefficient vector is $\bm{\gamma} = (\gamma_0, \gamma_1, \ldots, \gamma_{K+Q})^T$. 

The covariance function we employ in equations (\ref{eq:3.32})  - (\ref{eq:spa5}) is a special case of the Mat\'ern class of spatial covariance functions, for modeling the dependence between spatial observations [\Mycite{gelfand2010}]. For instance, a large value of $\rho$ suggests that countries that are relatively far from one another are still moderately correlated [\Mycite{hoeting2006}]. Note that while we consider GCDs, geographical distance measures on a global scale have always been a contentious issue [\Mycite{ward2018}]. We discuss some possible extensions to the spatial component in our framework in Section \ref{ssec:spadist}.
\vspace{-1.3em}
\section{Latent Health for the World}\label{sec:LHW}
We present results from our LACSH model (spatial causal LHFI in Section \ref{ssec:BPSA}), separately fitted to the countries' data using MML days and GGHE as the treatment variable. The posterior credible intervals of rankings of countries are provided in Appendix B, and the resulting LACSH rankings using the two different treatment variables are compared to each other and also to the HDI and SPI rankings on ladder plots in Appendix C.

For Bayesian inference, an adaptive MCMC algorithm [\Mycite{roberts2009}] was used to automatically tune all parameters on the $H$-level, and as well as $H_{anc}$ in the MCMC due to non-conjugacy and to improve convergence and mixing. All other parameters were sampled using Gibbs sampling. Specific sampling specifications are documented in Appendix A. For model results, we utilized roughly 100,000 post-burn-in MCMC samples from the posterior distribution. Standard diagnostics (e.g. trace plots in Table \ref{tab:coefB} and effective sample sizes in Appendix D) suggested that each parameter of the MCMC chain had reached its steady state. 
\vspace{-1em}
\subsection{Priors} \vspace{-0.8em}
We specify conjugate diffuse priors for most parameters. For each regression coefficient $a_j$, $\beta_k$, and $\gamma_l$, the variance parameter log($\sigma^2_H$), and the spatial correlation parameter log$(\phi)$, we specify a normal prior distribution with mean 0 and variance 100. The covariance matrix $\Sigma_Y$ is given an inverse-Wishart prior with $P + 2$ degrees of freedom, and an identity scale matrix. The diffuse prior for $\sigma^2_T$ is inverse gamma with shape = 1 and scale = 0.01. 
\vspace{-1em}
\subsection{Ranking of countries according to latent health, $H$} \vspace{-0.5em}
\subsubsection*{I. MML days} \vspace{-1.4em}
\begin{figure}[H]
	\includegraphics[width=\textwidth, height=0.595\textheight]{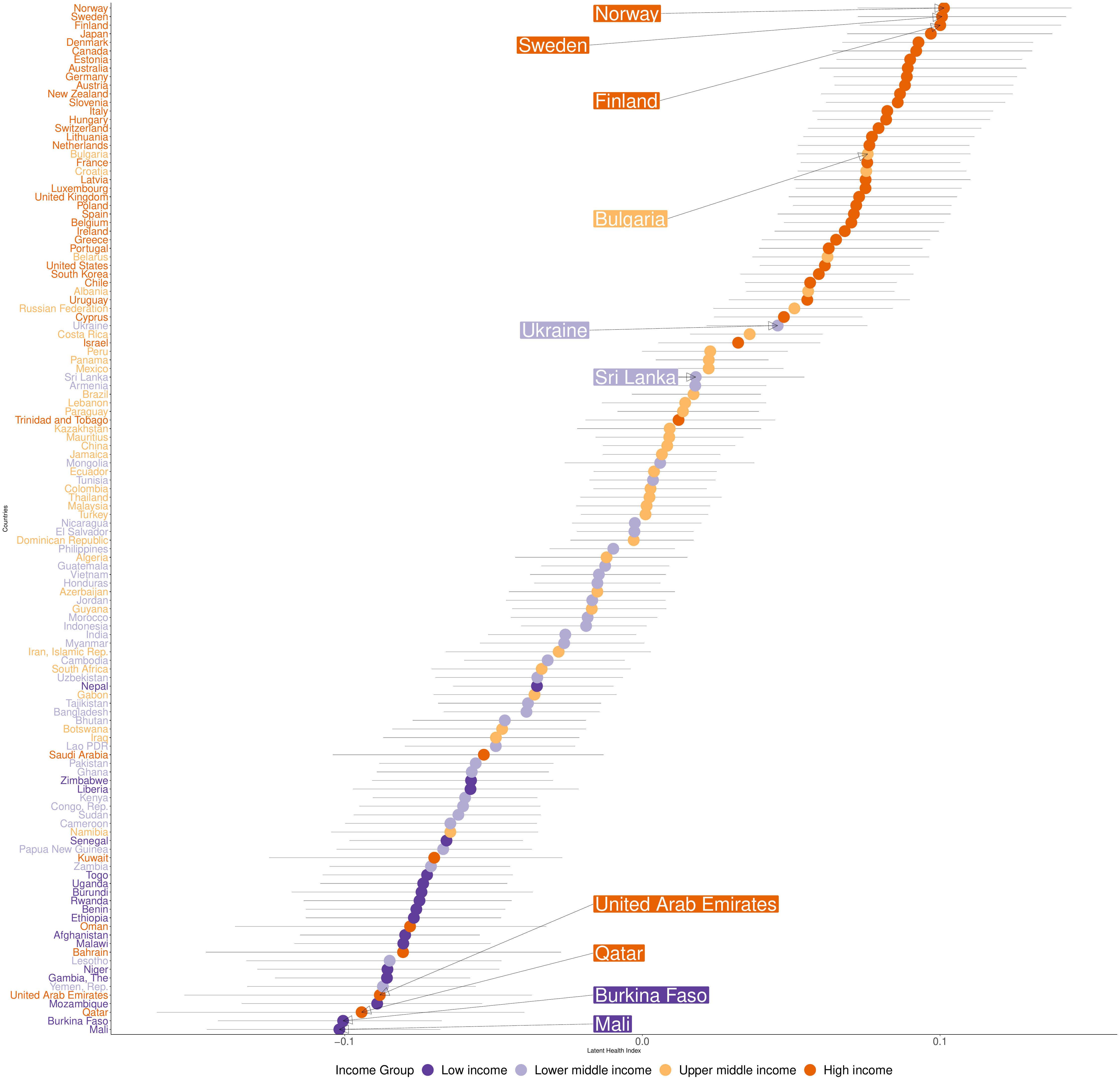}
	  \vspace{-2em}
	\caption{Latent health for 120 countries in 2016 color-coded by income group, with MML as treatment}	  \vspace{-2em}
	 \label{fig:HrankB1}
\end{figure}

\textcolor{white}{a} Figure \ref{fig:HrankB1} shows the country ranking based on the posterior medians of the $H$'s (colored dots) along with their corresponding 90\% credible intervals (gray bands). The figure highlights some countries that are ranked highest, lowest, or differently than its United Nations' (UN) designated income group\footnote{UN and the World Bank classify countries every year into four income groups based on their GNI per capita (current US\$).}. Exceptions include United Arab Emirates (UAE) and Qatar, which belong to UN's high income group, but are ranked low by our LACSH index when using MML as policy treatment, and also by the Social Progress Index [\Mycite{stern2020}]. Indeed, MML may be seen as a strong social progress indicator. Formal quantification of the uncertainty for our health parameter suggests that countries are not polarized into developed/developing countries or rich/poor countries; the lack of polarization aligns with the findings in \Mycite{rosling2019}. 

Nevertheless, our color-coding according to the designated income groups shows that the countries are generally ranked according to their income group. This suggests that the health of a country is highly correlated with the income group of the country. However, as will be discussed below, income is not necessarily the most important index to examine when considering the health of a country.

The posterior median and corresponding credible interval for the highest-ranked, lowest-ranked, and anchor country are shown in Table \ref{tab:coefB}.

\subsubsection*{II. GGHE} \vspace{-0.5em}
\begin{figure}[H]
\vspace{-0.5em}
	\includegraphics[width=\textwidth, height=0.595\textheight]{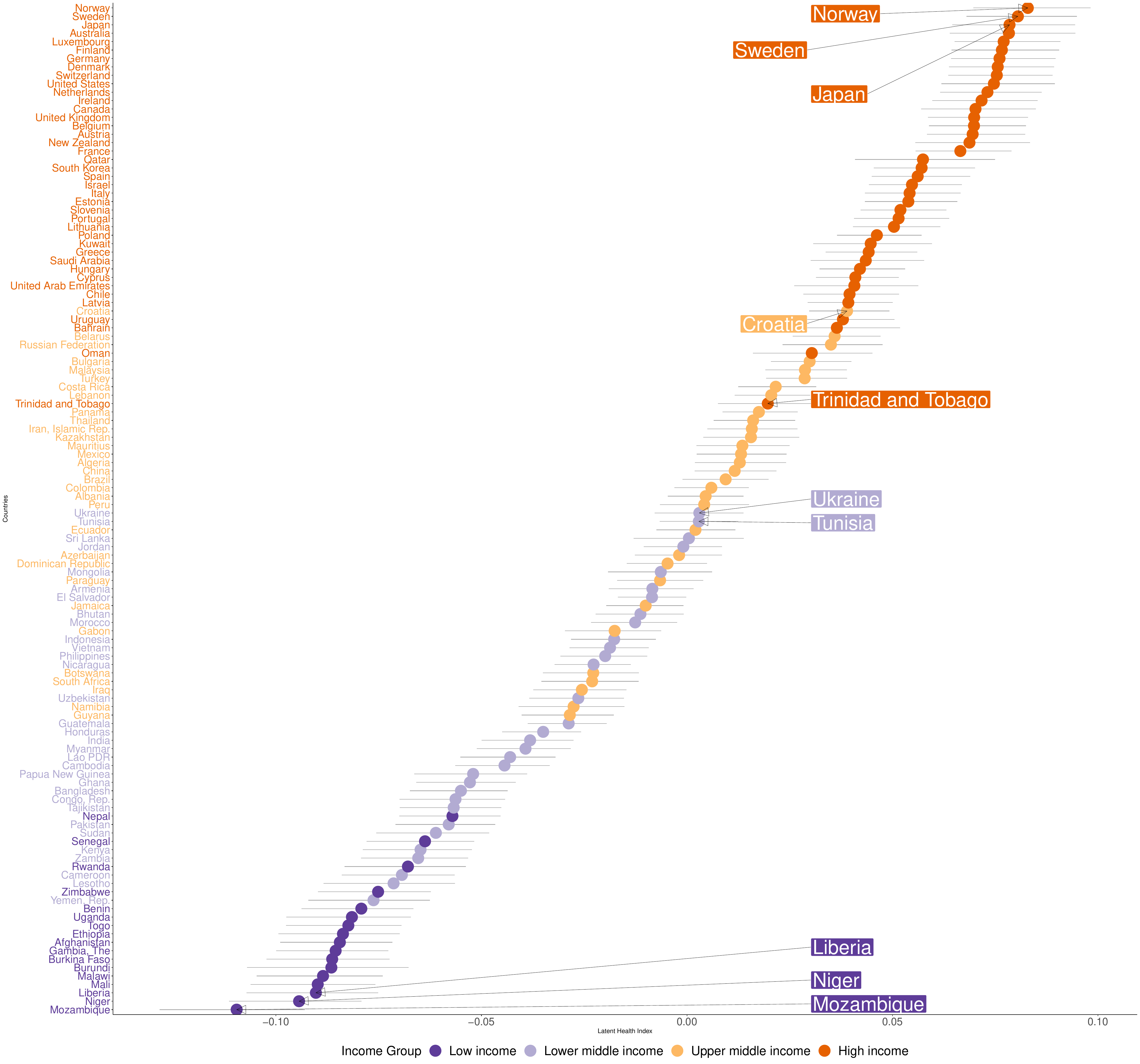} 
	\caption{Latent health for 120 countries in 2016 color-coded by income group, with GGHE as treatment}
	\label{fig:HrankB2}
  \vspace{-2em}
\end{figure}

The countries' health ranking using GGHE as the treatment variable is shown in Figure \ref{fig:HrankB2}. Compared to using MML as treatment, this ranking shows tighter credible intervals (narrower gray bands) and follows the UN's designated income groups more closely. This may be due to the high correlation between the GGHE treatment variable and many metrics and covariates (e.g. `\texttt{mean years of schooling}', `\texttt{GNI per capita}').

Some exceptions are Ukraine and Sri Lanka, which are classified as lower-middle income countries by the United Nations, but are ranked among the high and upper-middle income countries in our rankings in both models fitted using MML and GGHE as treatment. These results suggest that a country's health is not solely reflected by its income or wealth.
\vspace{-0.5em}
\subsection{Numerical results and implications}\label{ssec:coefB}
Posterior summaries for selected parameters are shown in Table \ref{tab:coefB}. The other model parameters are tabulated in Appendix B.

\subsubsection*{Nations' latent health, $H$}
Table \ref{tab:coefB} left panel shows the highest and lowest ranked $H_i$ from our LACSH model using MML as treatment, corresponding to Norway ($i =$ 84) and Mali ($i =$ 73). Based on the 90\% credible intervals, Figure \ref{fig:HrankB1} suggests that there are no stark differences from country to country. Nevertheless, we can consider potential groupings by examining the posterior probability of a positive difference in health between countries. For the top-ranked countries, the posterior probability for Norway to be in better health than Sweden is negligible at 0.53, for Sweden to be better than Finland is 0.51, whereas for Finland to be better health than Japan is 0.57, suggesting the four countries may be grouped together. Similar calculations of posterior probabilities can also be easily obtained for other countries. 

Some of the posterior summaries for our LACSH model using GGHE as treatment are tabulated in the right panel of Table \ref{tab:coefB}. Norway ($i =$ 84) is again ranked at the top, similar to using MML as treatment, followed by Sweden ($i =$ 104), Japan ($i =$ 57) and Australia ($i =$ 5). However, similar calculations of posterior probability of difference as above again suggest these countries may be grouped together. Mali and Mozambique remain in the bottom five of our ranking for results from both treatment variables. 

\subsubsection*{Health loadings, $\bm{a}$} \label{ssec:metric}
Insights into the associated strength and direction of relationship between metrics and health are available from the inference about the health loadings, $a_j$. These results demonstrate that our model-based approach does not require \textit{a priori} input on which metrics reflect `good' health, or which metrics are important to a country's latent health.

\begin{figure}[!h]
\includegraphics[width=0.85\textwidth,height=0.35\textheight]{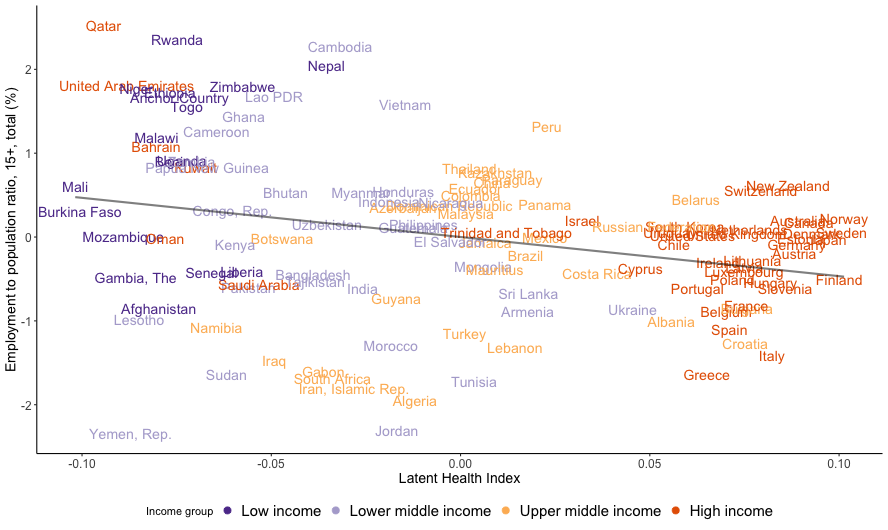} 
\caption{Plot of employment-to-population ratio vs. LACSH model posterior median of latent health, with MML as treatment, and with a least squares regression fit for visualization}
\label{fig:employ1}
\end{figure}

Table \ref{tab:coefB} left panel shows the results based on the marginal posterior medians using MML as treatment variable, for the three metrics which receive the highest positive impact from health, and one example of a negative loading, which will be discussed via Figure \ref{fig:employ1}. 

In decreasing order of effect size, the corresponding positively affected metrics are: `\texttt{population, ages 65 and older (\% of total)}'' ($j =$ 9), `\texttt{education index}' ($j =$ 1) and `\texttt{infant mortality rate (per 1,000 live births)}' (reversed-scale) ($j =$ 6). This suggests that, among the 15 metrics,  population health-themed and education variables receive the highest three loadings from the country's latent health. 

The right panel of Table \ref{tab:coefB} shows some of the results for GGHE as treatment. The posterior probability is 0.66 for health to have a bigger effect on `\texttt{GNI per capita}' ($j =$ 3) than `\texttt{infant mortality rate (per 1,000 live births)}' (reversed-scale) ($j =$ 6), and is 0.55 for `\texttt{infant mortality rate}'  than `\texttt{internet users(\% of population)}' ($j =$ 4), suggesting that the latent health of countries has similarly weighted effects on the three metrics receiving the highest positive impact. The different results between MML and GGHE suggest that which metrics are most affected by health depends on the treatment variable in the model. Other relevant discussions for GGHE as treatment are given in Appendix E. 

\FloatBarrier
\begin{table}[htb!]
	\setlength\tabcolsep{2pt}
\begin{tabular}{cccccc}
\toprule
Parameter & {MC\footnote{Markov chain}} & 5\% & Median & 95\%  \\
\midrule 
$H_{84\phantom{,} \text{(Norway)}}$ & \includegraphics[width=0.05\textwidth, height=0.03\textheight]{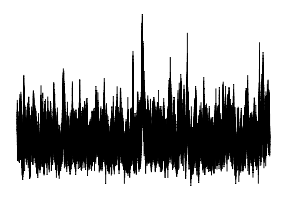} & 
0.07 & 0.10 & 0.14 \\
$H_{73\phantom{,} \text{(Mali)}}$ & \includegraphics[width=0.05\textwidth, height=0.03\textheight]{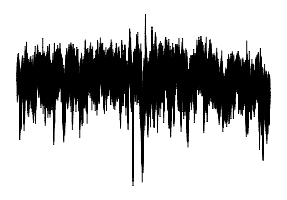} & 
-0.15 & -0.10 & -0.07 \\
$H_{anc}$ & \includegraphics[width=0.05\textwidth, height=0.03\textheight]{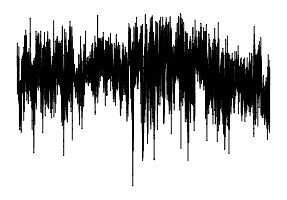} & -0.12 & -0.07 & -0.04 \\
$a_9$ & \includegraphics[width=0.05\textwidth, height=0.02\textheight]{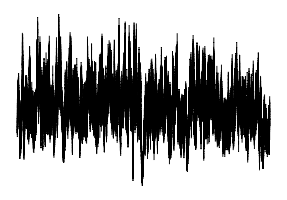} & 10.97 & 14.80 & 19.02 \\
$a_1$ & \includegraphics[width=0.05\textwidth, height=0.02\textheight]{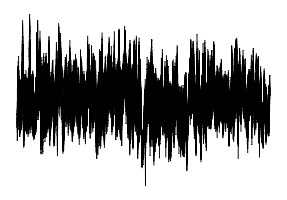} & 9.07 & 12.74 & 16.68 \\
$a_6$ & \includegraphics[width=0.05\textwidth, height=0.02\textheight]{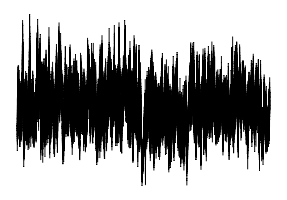} & 8.90 & 12.73 & 16.78 \\ 
$a_{2}$ & \includegraphics[width=0.05\textwidth, height=0.02\textheight]{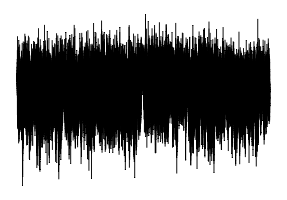} & -7.35 & -4.17 & -1.38 \\
$\beta_1$  & \includegraphics[width=0.05\textwidth, height=0.03\textheight]{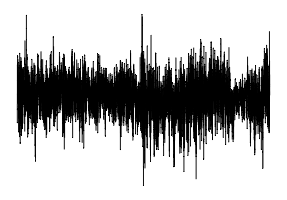}& -0.02 & -0.01 & 0.00 \\
	$\sigma_H^2$  & \includegraphics[width=0.05\textwidth, height=0.03\textheight]{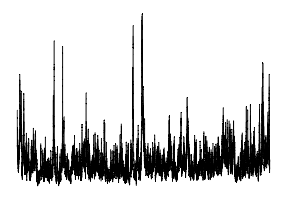} & 0.002 & 0.005 & 0.011 \\
$\phi$ & \includegraphics[width=0.05\textwidth, height=0.02\textheight]{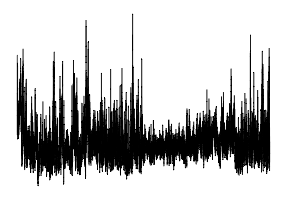} & 3.37 & 5.37 & 8.82 \\
$\gamma_1$ & \includegraphics[width=0.05\textwidth, height=0.02\textheight]{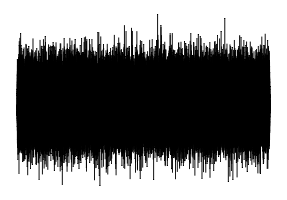} & -0.10 & 0.04 & 0.17 \\
\bottomrule
\end{tabular}
\hspace{0.5em}
\begin{tabular}{cccccc}
\toprule
Parameter & {MC} & 5\% & Median & 95\%  \\
\midrule 
$H_{84\phantom{,} \text{(Norway)}}$ & \includegraphics[width=0.05\textwidth, height=0.03\textheight]{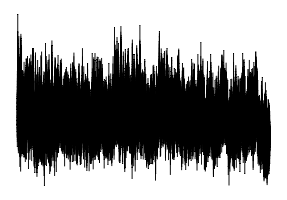} &  {0.07} & 0.08 & {0.10} \\ 
$H_{76\phantom{,} \text{(Mozambique)}}$ & \includegraphics[width=0.05\textwidth, height=0.03\textheight]{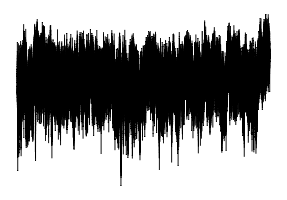} & 
{-0.13} & -0.11 & {-0.09} \\ 
$H_{anc}$ & \includegraphics[width=0.05\textwidth, height=0.03\textheight]{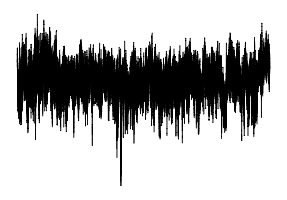} & {-0.11} & -0.09 & {-0.07} \\ 
$a_3$ & \includegraphics[width=0.05\textwidth, height=0.02\textheight]{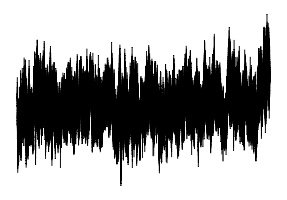} & 
 {15.43} & 17.81 & {20.23}  \\
$a_6$ & \includegraphics[width=0.05\textwidth, height=0.02\textheight]{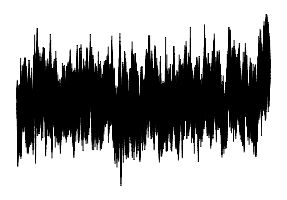} & 
{15.10} & 17.41 & {19.83} \\
$a_4$ & \includegraphics[width=0.05\textwidth, height=0.02\textheight]{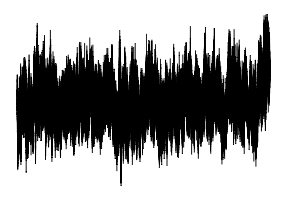} & 
{14.97} & 17.30 & {19.73} \\ 
$a_{15}$ & \includegraphics[width=0.05\textwidth, height=0.02\textheight]{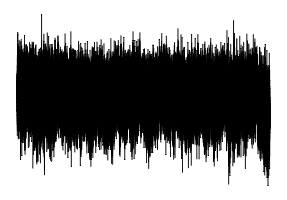} & 
{-13.64} & -10.86 & {-8.26} \\
$\beta_1$  & \includegraphics[width=0.05\textwidth, height=0.03\textheight]{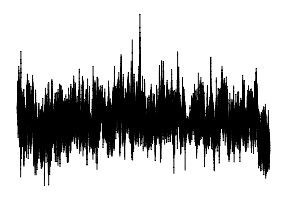}& {0.04} & 0.04 & {0.05}  \\
$\sigma_H^2$  & \includegraphics[width=0.05\textwidth, height=0.03\textheight]{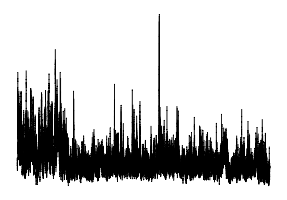} & {0.002} & 0.003 & {0.004} \\
$\phi$ & \includegraphics[width=0.05\textwidth, height=0.02\textheight]{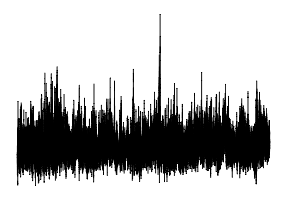} & {27.10} & 29.64 & {33.16} \\
$\gamma_1$ & \includegraphics[width=0.05\textwidth, height=0.02\textheight]{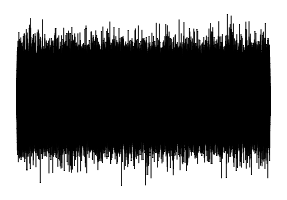} & {-0.04} & 0.11 & {0.26} \\
\bottomrule
\end{tabular}
\caption{Posterior summaries for selected LACSH model parameters, with MML and GGHE as treatment, respectively}
\label{tab:coefB}
\end{table}
\FloatBarrier

Figure \ref{fig:employ1} shows a weak negative relationship between the metric `\texttt{employment to population ratio}' and a country's latent health (90\% credible interval for $a_{2}$ is $(-7.35, -1.38)$). As we can see from the figure, countries with a high proportion of employment-to-population ratio are generally in the low-income group, and the ratio decreases with increasing income. This perhaps goes against the naive belief that a high employment ratio reflects a country's `good' health. There are several possible explanations for this result. One, wages are typically low in low-income countries, resulting in a higher proportion of the population having to work in order to secure a decent living wage. Two, lower-income countries relatively lack effective social safety nets and social protection systems for its population, resulting in a higher proportion of the population working for a longer period of time until (a possibly later) retirement age. Some exceptions are Qatar and UAE, which are high-income countries but also have high employment ratio due to a large proportion of its population being expatriate workers [\Mycite{parcero2017}].

Given the other metrics in the model, for MML as treatment, three metrics namely, `\texttt{population density}', `\texttt{unemployment rate}' (reversed-scale), and `\texttt{renewable energy consumption}' were found to have no substantial statistical relationship with a country's latent health. For GGHE as treatment, the metric `\texttt{proportion of seats held by women in national parliament}' also shows no statistical relationship with a country's latent health as opposed to `\texttt{renewable energy consumption}'. In fact, the model suggests there is a negative relationship between `\texttt{renewable energy consumption}' and a country's latent health. We provide further discussions in Appendix E.

\subsubsection*{Average dose-response function}
To examine the dose-response function which relates the inherent impact of varied levels of a policy treatment variable on the country's \textit{health}, we utilize the GPS formulation as proposed by \Mycite{Hirano2004}. The GPS approach allows us to evaluate a country's health outcome that corresponds to each specified value of the continuous treatment (i.e. MML days or GGHE). Note that the conditional expectation of the outcome as a function of the treatment $T$ and the GPS $R$ is $\beta(t, r) = E[H|T=t, R=r]$ for each specified treatment `dose' $t \in T$.

\begin{figure}[htb!]
  \includegraphics[width=0.95\textwidth]{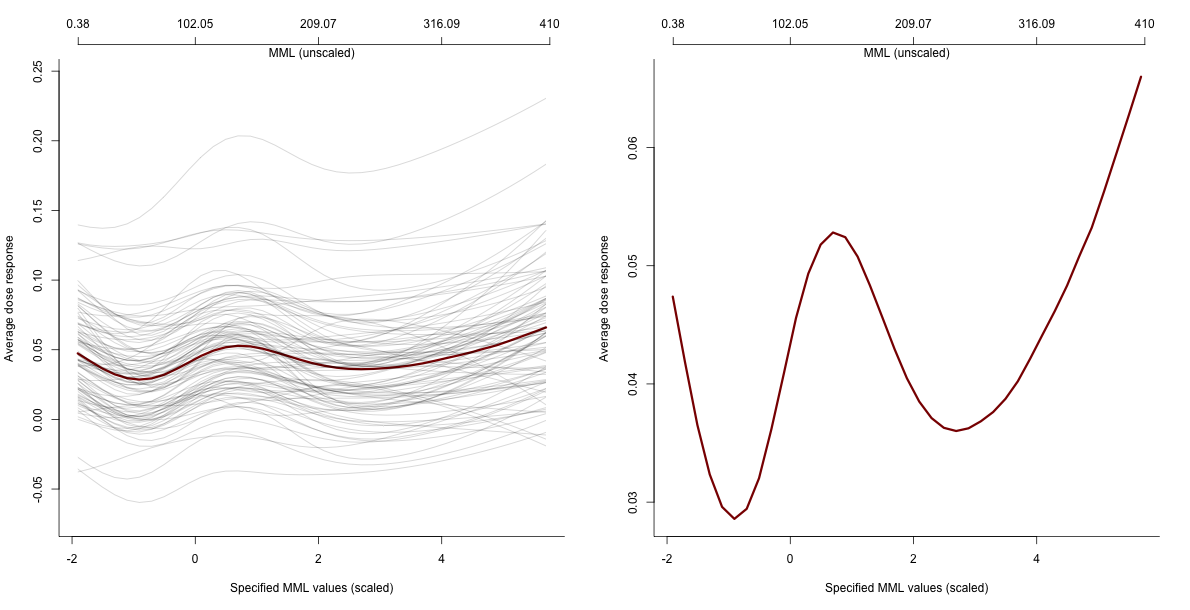}
      \caption{The LACSH model posterior median dose-response function (red), with MML as treatment, and with 100 thinned MCMC samples (gray) for visualization of uncertainty}
      \label{fig:DR_MML}
\end{figure}

\begin{figure}[H]
  \includegraphics[width=0.9\textwidth]{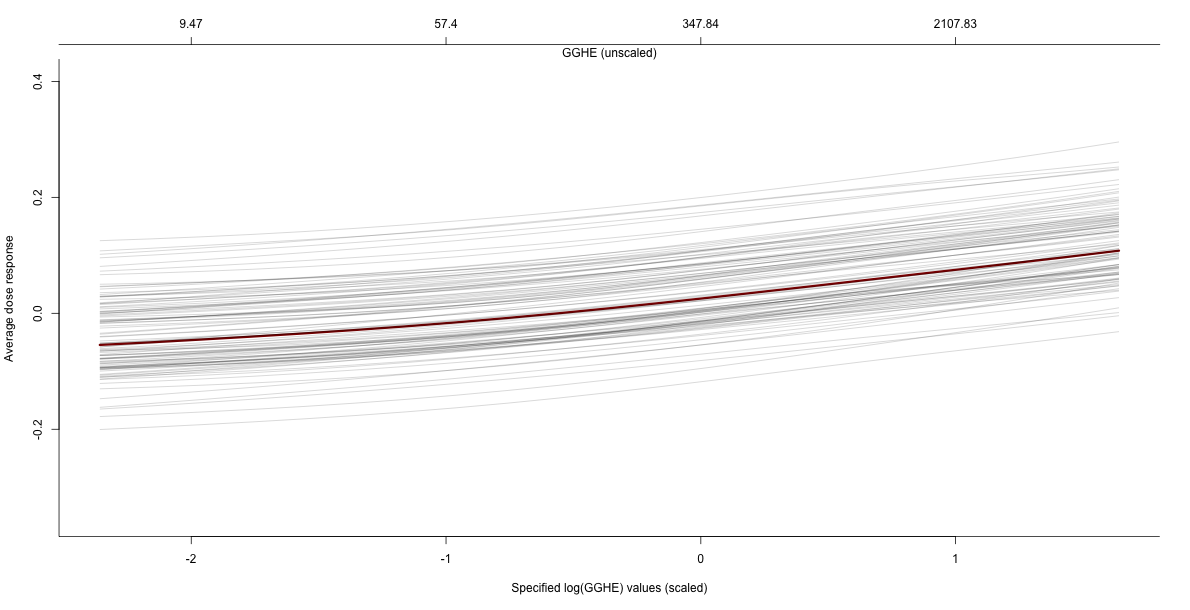}
    \caption{The LACSH model posterior median dose-response function (red), with GGHE as treatment, and with 100 thinned MCMC samples (gray) for visualization of uncertainty}
    \label{fig:DR_newtrt}
  \vspace{-2em}
\end{figure}

We can obtain an estimate of the entire dose-response function through estimating the average potential outcome at a given $t$. In particular, $\mu(t) = E[H_i(t)]$ is calculated similar to equation (\ref{eq:H_mean}) but substituting specific values of $t$ for $T$ in equations (\ref{eq:H_mean}) and (\ref{eq:trt2}). The entire dose-response function is then $\mu(t) = E[\beta\{t, r(t, \bm{Z_i}^*)\}]$, which is estimated by $\hat{\mu}(t) = (1/N)\sum_i \beta\{t, r(t, \bm{Z_i}^*)\}$. As $\hat{\mu}(t)$ depends on the parameters $\bm{\beta}$ and $\bm{\gamma}$, we examine the posterior median of $\hat{\mu}(t)$ at a given $t$ (red curves in Figures \ref{fig:DR_MML} and \ref{fig:DR_newtrt}).

\Mycite{Hirano2004} assert that the conditional expectation of the outcome as a function of the treatment level $T$ and the GPS $R$ ($\beta(t,r)$) does not have a causal interpretation, but that $\mu(t)$ which corresponds to the dose-response function for treatment level $t$, when compared to another value of $t'$, does have a causal interpretation.

Many authors in the GPS literature use quadratic and interaction terms of the GPS and treatment variable in its conditional expectation of the outcome. We examine the need for the quadratic terms, by comparing both models with and without the terms in equation (\ref{eq:H_mean}) using the Watanabe-Akaike information criterion (WAIC), also known as the widely applicable information criterion, which is useful for models with a hierarchical structure and in which the number of parameters grows with the sample size [\Mycite{watanabe2010}; \Mycite{gelman2014}; \Mycite{stat2020}]. We find that quadratic terms are needed for GGHE but not MML (WAIC values 3032.35 (quadratic) and 3026.84 for MML, and 2950.37 (quadratic) and 3742.99 for GGHE). To discuss the comparison between MML and GGHE, we keep the quadratic terms.

Figures \ref{fig:DR_MML} and \ref{fig:DR_newtrt} show the posterior average dose-response function using thinned posterior samples (for visualization purposes) from our LACSH model (eq. (\ref{eq:eq2}) - (\ref{eq:trt2})).

The left panel in Figure \ref{fig:DR_MML} shows gray curves as the posterior dose-response based on a thinned MCMC sample; the red curve indicates the posterior median based on approximately 100,000 MCMC scans. The curves show a very weak increase (with a slight blip) in the average dose-response as the number of MML days increases. The right panel, the curve of posterior median zoomed-in vertically, shows an increasing dose-response when MML days range over 53-145 and 236-410. Upon further inspection, the double `dip' may be due to the lack of data around certain ranges of MML days with clusters of countries having similar MML values.

\vspace{-0.5em}

The average dose-response function shown in Figure \ref{fig:DR_newtrt} suggests that increasing the level of health expenditure (GGHE) monotonically leads to an increased level of the country's health. The figure also presents strong evidence of a causal phenomenon, in that all posterior samples in the figure are monotone.

\subsubsection*{Assessing covariate balance under the generalized propensity scores framework} \label{ssec:PS}
In Rubin's approach of causal inference, the balance between treatment and control groups with respect to the pre-treatment covariates is a crucial assumption [\Mycite{imbens2015}]. We introduce a novel technique to visually assess covariate balance in the GPS framework. Generally, testing for covariate balance with a continuous treatment variable is not straightforward, and we consider the approaches by both \Mycite{Hirano2004} and \Mycite{imai2004} (as noted in \Mycite{kluve2012}) to test for:
\vspace{-0.1em}
$$ \bm{Z_i^*} \perp \mathbb{1}\{T_i=t\} \ | \ r(t, \bm{Z_i^*}) $$

\noindent where the GPS $r(\cdot)$ is evaluated at different specified values of $t$ for the continuous treatment variable. Specifically, we divide the sorted data of $\{T_1, \ldots, T_N\}$ into moving blocks of 20 observations, overlapping 10 observations between neighboring blocks. (This results in 11 blocks for $N=$120 observations.) The GPS is then evaluated at the median of each block.

As the GPS is a dimension reduction tool to control for what is usually a large number of covariates, we investigate covariate balance collectively instead of what is done in the literature, being on each individual covariate. Specifically, we represent the covariates by its first principal component $f(\cdot)$ when evaluating
\vspace{-0.05em}
\begin{eqnarray} f(\bm{Z_i^*}) \perp \mathbb{1}\{T_i=t\} \ | \ r(t, \bm{Z_i^*})
\label{eq:indp} 
\end{eqnarray}

\noindent where $t$ is the block median, 
\begin{center}
$\mathbb{1}\{T_i=t\} = 
\left\{ \hspace{-1.2em}
\begin{array}{ll}
 &1 \ \text{if country } i \text{ is in the current block of 20 countries} \\
 &0 \ \text{otherwise} 
  \end{array}
\right .
$
\end{center}
\vspace{-0.8em}
\noindent and $r(\cdot)$ is computed as $r(t; u_i, v) = \frac{1}{v \sqrt{2\pi}} \text{exp}(-\frac{(t - u_i)^2}{2v^2})$ where $u_i$ and $v$ are based on a standalone frequentist linear multiple regression of the $T$ data on the $\bm{Z^*}$ data, such that $u_i$ is the fitted value for the $i$th country, and $v$ is the fitted residual standard error. 

\begin{figure}[H]
    \vspace{-0.5em}
	\includegraphics[width=0.9\textwidth, height=0.28\textheight]{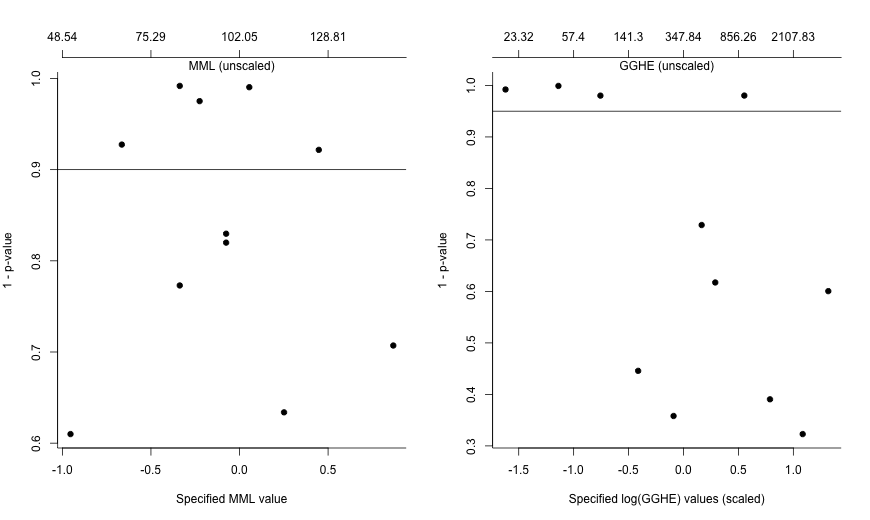}
	\caption{Plot of p-values (reversed-scale) for assessing covariate balance, with MML as treatment (left panel), and with GGHE as treatment (right panel).}
	\label{fig:PSass} 
\end{figure}

To evaluate if equation (\ref{eq:indp}) holds, first, for each block of 20 countries, we run a frequentist logistic regression of $\mathbb{1}\{T_i = t\}$ on $f$ and $r(\cdot)$ and record the $p$-value of the slope for $f$. A $p$-value exceeding, say, 0.1, suggests that the covariates, given the GPS, are not significantly related to the treatment variable at and around that median value. Because the main objective of the GPS methodology is to remove any potential biases introduced by the covariates, we consider the set of 11 $p$-values collectively, whereby covariate balance is deemed adequate if most of the 11 $p$-values satisfy 0.9 $>$ 1-$p$. In the case of MML, left panel of Figure \ref{fig:PSass} shows there are 5 out of 11 blocks that are above the 0.9 threshold, but if we used 0.95 then 8 out of 11 blocks show adequate covariate balance. The right panel of Figure \ref{fig:PSass} shows the same blocks that contribute to the overall imbalance whether we use 0.9 or 0.95 as the threshold. 

\vspace{-0.7em}
 \vspace{-1.5em}

In addition to `blocking on the (generalized propensity) score' to assess covariate balance [\Mycite{imai2004}], our novel approach described here also allows us to identify the ranges of treatment values that may be the sources of any overall imbalance. The observations that fall under those ranges may be further investigated for improving overall covariate imbalance, if necessary. In the case of MML (left panel of Figure \ref{fig:PSass}), there are no consistent patterns of covariate imbalance. In contrast, in the case of GGHE (right panel of Figure \ref{fig:PSass}), the lower range of GGHE values (approximately 0-90) consistently show covariate imbalance. To further investigate this, we removed countries in those ranges and re-ran our model on the subsample of countries, which removed the consistent pattern of imbalance. More details are in Appendix F.
\vspace{-1em}
\subsubsection*{Inverse decay parameter $\phi$ and spatial correlation function $\rho(d, \phi)$}
\vspace{-1em}
\begin{figure}[H]
\includegraphics[width=0.9\textwidth, height=0.28\textheight]{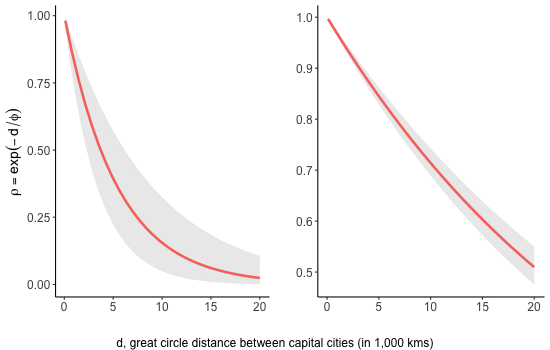}
\caption{Spatial correlation based on the posterior of $\rho = exp(-\frac{d}{\phi})$ as a function of $d$, under the LACSH model with MML as treatment (left panel), and with GGHE as treatment (right panel).}
\label{fig:rho_covar1}
\end{figure}

Figure \ref{fig:rho_covar1} shows the posterior median of the spatial correlation function $\rho$ evaluated at a given $d$ between capital cities, under the LACSH model with MML and GGHE as treatment, respectively. Both panels show decreasing spatial correlation between the countries as distance increases. The model with MML has a sharper decrease but with more uncertainty (wider credible band) compared to GGHE. For the two countries whose capitals are furthest apart (Spain and New Zealand), our models yield noticeable spatial correlation (90\% credible intervals of $(0.02, 0.10)$ and $(0.51, 0.55)$) with MML and GGHE as treatment, respectively. These results suggest that the LACSH models would be inadequate if spatial dependence were not accounted for.

\section{Some technical details}\label{sec:discussion}
\subsection{Identifiability}\label{ssec:ident}
Recall that at the metric level (Y-level), $a_j$ is the population-level loading of a country's health on its $j$th metric. However, similar to the discussions by \Mycite{chiu2011a} and \Mycite{martin2002}, modeling health $H_i$ as a random effect leads to an unidentifiable $\bm{a}$ vector unless constraints are imposed. 

The constraint we have utilized in our models is a TMVN distribution on the $H$-level so that the health of an anchor country is negative (or positive, if desired). To decide on the anchor country, we conducted a pilot run of the base LHFI model in Section \ref{ssec:LHFI} but without any anchor, then selected a low-income group country (Burundi, in this case) on the extreme end of the $H$-scale as the anchor in all subsequent formal models.

As the ranking of a country is relative to the others', the constraint restricts the anchor country's health in the negative space and imposes this fixed scale on all other countries. This constraint solves the parameter identifiability issue along with aiding the interpretation of $H_i$, as it encodes in the model the fact that a higher value of $H_i$ should be interpreted as a higher level of health, not lower. 

We have explored alternative constraints, including a `soft anchor' (fixing mean and variance for $H_{anc}$), a `hard anchor' (fixing $H_{anc}$ to a constant), and transposing the $H$-scale manually post-MCMC sampling (so the scale of $H$ aligns with increasing value of $H$ corresponding to increasing levels of health). While our `truncated anchor' leads to additional computational burden, we believe this constraint to be the most desirable as it results in the most flexible approach. 

\subsection{Spatial distances}\label{ssec:spadist}
Section \ref{ssec:coefB} shows that there is considerable spatial correlation among countries, so that not accounting for it would result in model inadequacy. To limit the amount of additional model complexity in our LACSH model, spatial correlation between countries was modeled through the simplest Mat\'ern covariance function in the form of an exponential decay over great circle distances between capital cities of countries. Due to the earth's spherical nature, Euclidean distances may be inappropriate on a global scale [\Mycite{banerjee2005}], and therefore to define neighbors,  \Mycite{gleditsch2001} also use a minimum distance between countries based on country borders of up to a certain distance. When we extend the LACSH model to more complex forms in order to formally incorporate dependencies jointly across time (2010--2016) and space (see Section \ref{sec:future}), we may explore this minimum distance measure, as well as alternative covariance functions. 

\subsection{Data transformation and missing values}\label{ssec:missing}
Higher unemployment rate is generally regarded as bad for societal health [\Mycite{wulfgramm2014}; \Mycite{helliwell2014}]. For this reason, the metric for the \texttt{`infant mortality rate'} had been linearly transformed prior to modeling so that higher values reflect better societal health. The same transformation was also applied to the metric for \texttt{`unemployment rate'}. In practice, the modeler needs not to carry out this transformation, because the fitted model can be used to distinguish the strength and direction of the relationship between latent health and metrics, as reflected by the signs of the health loadings $a_j$ on metrics. For instance, the results from both of our treatment variables suggest that higher values of latent health are associated with higher values of the (reversed-scale) infant mortality rate metric (equivalent to low infant mortality rate in the country before the data transformation). In other words, it is inferred that countries with better latent health have low infant mortality rate, conditioned on the set of metrics and covariates that are included in our model. Other visualizations and summaries of the posterior distribution for the negative health effect, $a_j$, are included in Appendix E in the supplementary material. 

Note that the selection of variables to be included as metrics and covariates in our model was largely based on the availability of data. Initial model fits included more covariates; however, due to collinearity, past-year metrics with a correlation higher than the arbitrary threshold of 0.8 with other covariates/metrics were removed sequentially from modeling. While this paper focuses on the development of methodology, when applying the methodology in practice, the covariates and metrics could be specified by the modeler more according to their domain knowledge and less to data availability. In either case, the issue of missing data may require special attention. 

In this paper, we considered two continuous `policy treatment' variables separately in our models. In particular, for MML days, the data obtained from the World Bank only include MML data every other year. Therefore, data for the years 2010, 2012, and 2014 are considered missing. For those years, only countries with the same values for the years before and after the missing year entered our model. To further reduce data missingness in each year, OECD data were used for some OECD countries when MML was missing from the World Bank data, although we note that the two organizations have slightly different definitions of maternity leave. For example, according to the World Bank, Sweden has zero MML days based on its definition. However, the OECD and other sources suggest that this may not be an accurate representation of their maternity policy. (As such, future iterations of our work will consider non-World Bank definitions.) 

In our work, given the right-skewed nature of GGHE, we log-transformed the values before standardization. MML days also appeared skewed, but the data appeared in clusters, thus showing large unobserved ranges, and various transformations did not improve the data distribution. For this reason, we had kept the variable untransformed other than standardizing it to have mean zero and unit variance. In future extensions of the current work, we may consider other approaches such as rank likelihood estimation when using treatment variables that are distributed in clusters. All transformations including those in Table \ref{tab:variables} are intended to adhere to our normality assumptions.

Finally, even if data exist in published records, it is recognized that such data collected on the country-level by various world organizations may have been derived from different and unpublished imputation techniques. Of course, the quality of the data would depend on the actual imputation techniques employed. Moreover, one may not rule out the possibility that data or official statistics reported by certain countries may have been fabricated. Although these disadvantages could reduce the accountability of our modeling results, overcoming such data-related challenges is beyond the scope of our paper.

\section{Simulation studies}\label{sec:simStudies} 
To validate our models, we ran two simulations. The first is a cross validation study to recover missing metric $Y$ values. The second study assesses the ability to recover true parameter values. Detailed results of the simulations are shown in Appendix G. 


In the cross validation study,  we use the original data in the paper where our metric $\bm{Y}$ is an $N \times Q$ matrix, where $N = 120$ and $Q = 15$. We randomly select 100 values (about 5\%) from the original $N \times Q$ values of $y_{ij}$ and discard them. In the inference, the 100 missing values are treated as model parameters and estimated alongside all the other parameters. We repeat this process 10 times, each time resulting in a different set of missing values. For each set of 100, we record the number of withheld $y_{ij}$'s that are captured by the 95\% posterior predictive intervals. We find that among the 10 sets, the lowest capture rate for missing $y_{ij}$'s is 90\% (90/100) and the highest is 98\% (98/100).

In the second simulation study, we generate fictitious data based on a version of our model. We fix the parameter values on the $T$-, $H$- and $Y$-levels and generate new $T$'s, $H$'s and $Y$'s, each time the dimension for $\bm{Y}$ is $N^* \times Q^*$, where $N^* = 10$ and $Q^* = 3$. We set $\sigma_H = \sigma_T = 1$; $\phi = 5$; $\gamma = 0.5$; $\bm{a} = (0.14, 0.12, 0.08)$; $\Sigma_Y$ is fixed with a once-randomly generated $3 \times 3$ Wishart-distributed matrix with degrees of freedom 3. Three sets of $\bm{\beta}$ values were used --- $\{(2,3,3), (2,-3,-3), (2, -3, \frac{(-\beta_0 + \beta_1 \bar{T})}{\bar{R}})\}$, where $\bar{T}$ and $\bar{R}$ are medians of $T$ and $R(\cdot)$ values from the generated $T$'s. We fit our LACSH model and compute the 95\% credible interval for each parameter. We repeat the entire process 50 times for each of the three sets of $\bm{\beta}$ values. We find that the capture rate ranges from 70\% (35/50) to 100\% (50/50).

\section{Discussion and future work}\label{sec:future}
In this paper, we developed a LACSH model-based approach that models spatially correlated latent traits alongside the evaluation of policy effects through the integration of LHFI with a causal modeling framework. Through rigorous handling of quantitative treatment variables and the uncertainty quantification that directly results from the integrated LACSH modeling, we have demonstrated that our novel unified framework and visual assessment of covariate balance can be valuable to evidence-based social science with causal implications. Our methodology has been validated through simulations discussed in the previous section.

As mentioned in Section \ref{ssec:BPSA}, to facilitate formal causal inference, our LACSH approach incorporates the Bayesian extension of the generalized propensity score in the spatial LHFI framework. In addition, we intend to consider the alternative framework of Pearl's causal diagram approach, which could reveal if, by controlling for certain variables, we have unintentionally opened some `backdoor paths' in the causal diagram which would result in spurious correlation [\Mycite{pearl2009}]. In the literature, backdoor paths are any non-causal paths between the treatment and outcome variables in the causal diagram. As such, Pearl's approach might prompt us to explore controlling for a different set of variables and potentially lead to a different scientific conclusion. 

In addition to a spatial correlation function to account for dependencies using great circle distances between countries, one might also explore the use of clustering to account for potential cultural or political ideology of countries, but that is beyond the scope of our current paper. Note that other types of distance measures and neighborhood structures may also be explored, although island countries that do not share borders (e.g. Australia, New Zealand, and Sri Lanka) should be handled with care. Regarding temporal data, currently, metrics from 2016 are hierarchically regressed on the treatment from 2015 and the GPS, which depends on temporally averaged covariates and metrics from 2010--2014. A substantively more complex model would be required to formally model temporal correlation in any of $Y, H, T \text{ and } X,$ as an extension to this paper. This would result in a spatio-temporal hierarchical causal model. We anticipate that careful consideration of separability (or otherwise) between space and time will be required.

Also of interest may be formal inference that allows us to identify which metrics are crucial in reflecting the health of a country. To do so, we would consider modeling the health loadings as proportions that sum to 1, resulting in a type of variable/model selection framework. The implication of such a parameterization is the reduction in any modeler-induced selection bias due to choosing variables that are \textit{a priori} perceived as being important in reflecting a country's health.

Lastly, existing causal frameworks that jointly handle multiple treatment variables are not readily adaptable to our LACSH framework. The extended rank likelihood approach by \Mycite{kuh2022} may be further explored for this purpose.

\begin{acks}[Acknowledgements]
This research has been supported by an IBISWorld philanthropic donation by the late Phil Ruthven to the Australian National University in the form of research funds awarded to GS Chiu. We thank the handling editor and reviewers for their valuable input. We also thank Beatrix Jones, Bruce Chapman, Carolyn Huston, Corwin Zigler, Paul Gustafson, Peter Mueller, Tim Higgins, and the attendees of Bayes on the Beach 2017 and Joint Statistical Meeting 2018 for stimulating discussions and constructive comments on the topic. 
The authors acknowledge William \& Mary Research Computing (\url{https://www.wm.edu/it/rc}) for providing computational resources and technical support that have contributed to the results reported within this paper.
\nocite{horrace2005} \nocite{hoff2009} \nocite{hlavac2022}
\end{acks}

\appendix
\counterwithin{figure}{section}
	\section{MCMC algorithm}\label{appendix:MCMC}
	All model inference is done through MCMC sampling. All parameters are updated via Gibbs sampling except for some of the parameters on the $H$-level, because equations (4.2) and (4.4) are a combination of $N - 1$ multivariate normals and a truncated normal: 

	\begin{enumerate}
		\item Sample $H_{i|-i}$ from its N($M, V$) full conditional distribution where: 
		$$ V = \left( \boldsymbol{a}^T \Sigma_{Y}^{-1} \boldsymbol{a} + D^{-1} \right) $$
		$$ M = V^{-1}\left( \boldsymbol{a}^T\Sigma_{Y}\boldsymbol{y_i} + D^{-1} m_i \right)$$
		$$ D = \Sigma_{H[i,i]} - \Sigma_{H[i,-i]}^T \Sigma_{H[-i,-i]}^{-1} \Sigma_{H[-i,i]} $$ 
		$$ m_i = \mu_i + \Sigma_{H[i,-i]}^T \Sigma_{H[-i,-i]}^{-1} (H_{[-i]} - \mu_{-i})$$
		$$ \bm{\mu} = \mathbf{(1,T,T^2,R,R^2,T R)}\boldsymbol{\beta}$$
		$$\text{for } i = 1, \ldots, N; i \neq \text{anc} $$ 
		
		\item Sample $\bm{a} = (a_1, \ldots, a_P)^T$ from its MVN($\bm{M}$, $\bm{V}$) full conditional distribution where:
		$$ {\bm{V}} = \left( \sum_{i = 1}^{N} H_i^2\Sigma_{Y}^{-1} + 100 \bm{I_P} \right)$$
		$$ \bm{M} =  \bm{V}^{-1}  \left( \Sigma_{Y}^{-1} \mathbf{Y}^T \bm{H} \right) $$
 	         where $\bm{I_P}$ is a $P \times P$ identity matrix and $P = 15$.
		
		\item Sample $\Sigma_Y$ from its Inv-Wishart($\nu_n$, $\bm{S_n}$) full conditional distribution where:
		$$\nu_n = \nu_0 + N $$
		$$ \bm{S_n} = (\mathbf{Y} - \bm{H}\boldsymbol{a}^T)^T(\mathbf{Y} - \bm{H}\boldsymbol{a}^T)+ \bm{I_P} $$  
		where $ \nu_0 = P + 2$, $N = 120$ and $\bm{I_P}$ is as described above. 
		
		\item Sample $\sigma^2_T$ from its Inverse-Gamma($\alpha_n$, $\beta_n$) full conditional distribution where: 
		$$ \alpha_n = N/2 + 1 $$
		$$ \beta_n = \sum_{i=1}^{N} D_i^2/2 + 0.01$$
		$$D_i = T_i - \bm{Z_i}^* \bm{\gamma}$$

		\item Sample $\bm{\gamma} = (\gamma_1, \ldots, \gamma_9)^T$ from MVN($\bm{M}$, $\bm{V}$) where:
 
	$$ {\bm{V}} = \left(\sigma^{-2}_T\bm{Z^{*T}}\bm{Z^*} + 100\bm{I_{(1+K+Q)}} \right)$$
	$$ {\bm{M}} = \sigma^{-2}_T{\bm{V}}^{-1}\bm{Z^{*T}} \bm{T} $$

\noindent where $\bm{I_{(1+K+Q)}}$ is a $(1+K+Q) \times (1+K+Q)$ identity matrix and $(1+K+Q) = 10$. This MVN distribution is proportional to $\prod^n_{i = 1} P(T_i | \gamma, \sigma^2_T) P(\gamma) $ which is not a full conditional for $\bm{\gamma}$, in order to `cut the feedback' [McCandless et al. (2010); Zigler et al. (2013)]. This approximate conditional for $\bm{\gamma}$ is then used as the posterior predictive on the $H$-level. Note that this approximation ignores the ($H$-level) contribution from the (country's health) outcome, thus cutting the feedback.\\[0.15em]
 
\item Sample the $H$-level parameters ($\bm{\beta}^\star, \text{log}(\sigma^{2\star}_H), \text{log}(\phi^\star), H_{i=anc}^\star$) as a vector from the proposal distribution $Q_s(\bm{u}, \cdot)$ as set out below. In particular, the Metropolis algorithm is performed with a scan-specific proposal distribution; for $s \leq 200$, take $Q_s(\bm{u}, \cdot) = \text{MVN}(\bm{u}, (0.1)^2\bm{I}_d/d)$, whereas for $s > 200$, take $Q_s(\bm{u}, \cdot) = (0.9)\text{MVN}(\bm{u}, v^2\Sigma_s/d) + (0.1)\text{MVN}(\bm{u}, (0.1)^2\bm{I}_d/d)$ where $\bm{u} = $ our parameter vector in the previous MCMC iteration; $d = $ dimension of our target distribution; $v = 2.38$ for MML and $v = 5$ for GGHE. The different values for $v$ were adapted from Roberts and Rosenthal (2009) to improve slow mixing. \\

\noindent We define the ratio of densities as ${\displaystyle \kappa = \frac{p(\bm{\beta}^\star, \text{log}(\sigma^{2\star}_H), \text{log}(\phi^\star), H_{i=anc}^\star | \bm{R}, \bm{ T}, H_{i\neq anc})}{p(\bm{\beta}, \text{log}(\sigma^2_H), \text{log}(\phi), H_{i=anc} | \bm{R}, \bm{ T}, H_{i \neq anc})} }$ and accept $ (\bm{\beta}^\star, \text{log}(\sigma^{2\star}_H), \text{log}(\phi^\star), H_{i=anc}^\star) \text{ jointly with probability } \kappa \land 1. $
\end{enumerate}

\newpage
\section{Results for LACSH models}
\subsection{Posterior summaries for LACSH model, MML as treatment variable}\label{ssec:appC1}  
	\vspace{-0.5em}
	\begin{table}[htbp!]
     \caption{Posterior median of parameters using MML as treatment variable, and associated 90\% credible intervals} \vspace{0.1em}
	\label{tab:paramB1}
	\centering
	\footnotesize
	\resizebox{0.65\textwidth}{!}{\normalsize
		\begin{tabular}{rrrr}
		\toprule
		$a_j$ & 5\% & 50\% & 95\% \\
		\midrule
           	$a_{1}$ & 9.07 & 12.74 & 16.68 \\ 
 	        $a_{2}$ & -7.35 & -4.17 & -1.38 \\ 
        		$a_{3}$ & 6.39 & 10.20 & 14.15 \\ 
	        $a_{4}$ & 7.31 & 11.02 & 14.94 \\ 
       		$a_{5}$ & 8.53 & 12.35 & 16.39 \\ 
        $a_{6}$ & 8.90 & 12.73 & 16.78 \\ 
        $a_{7}$ & -3.17 & -0.11 & 2.93 \\ 
        $a_{8}$ & 8.06 & 11.58 & 15.42 \\ 
        $a_{9}$ & 10.97 & 14.80 & 19.02 \\ 
        $a_{10}$ & 4.66 & 8.14 & 11.90 \\ 
        $a_{11}$ & 2.02 & 4.83 & 8.14 \\ 
        $a_{12}$ & -4.47 & -1.44 & 1.46 \\ 
        $a_{13}$ & 7.40 & 10.88 & 15.00 \\ 
        $a_{14}$ & 6.84 & 9.83 & 13.34 \\ 
        $a_{15}$ & -7.66 & -3.59 & 0.53 \\ 
	\midrule 
 	$\sigma_H^2$  & 0.00 & 0.01 & 0.01  \\
	$\phi$ & 3.37 & 5.37 & 8.82 \\ 
	\midrule
	$\Sigma_{Y\{2,12\}}$ \tablefootnote{Only the top five in magnitude of the posterior median are presented.} & 0.68 & 0.53 & 0.68  \\ 
	$\Sigma_{Y\{3,15\}}$ & -0.47 & -0.69 & -0.47  \\ 
	$\Sigma_{Y\{10,15\}}$ & -0.40 & -0.61 & -0.40  \\ 
	$\Sigma_{Y\{3,4\}}$ & 0.40 & 0.22 & 0.40 \\ 
	$\Sigma_{Y\{4,15\}}$ & -0.40 & -0.62 & -0.40 \\ 
	\bottomrule
			\end{tabular}
		\begin{tabular}{rrrrc}
			\toprule 
			$\beta_k$ & 5\% & 50\% & 95\% & p($\beta > 0 \ |$ data) \\
			\midrule
  			     $\beta_{0}$ & -0.01 & 0.04 & 0.10 & 0.89 \\ 
 			     $\beta_{1}$ & -0.02 & -0.01 & 0.00 & 0.09 \\ 
       			     $\beta_{2}$ & -0.00 & 0.00 & 0.00 & 0.94 \\ 
    			    $\beta_{3}$ & -0.20 & -0.01 & 0.18 & 0.47 \\ 
  			     $\beta_{4}$ & -0.24 & 0.07 & 0.41 & 0.64 \\ 
        			    $\beta_{5}$ & 0.03 & 0.08 & 0.12 & 0.99 \\  
			\midrule
			   $\gamma_{(1+K+Q)}$ & 5\% & 50\% & 95\% &  p($\gamma > 0 \ |$  data) \\
			\midrule
  			    $\gamma_{0}$ & -0.11 & 0.00 & 0.11 & 0.50 \\ 
			      $\gamma_{1}$ & -0.10 & 0.04 & 0.17 & 0.68 \\ 
			      $\gamma_{2}$ & -0.21 & -0.08 & 0.05 & 0.15 \\ 
			      $\gamma_{3}$ & -0.11 & 0.02 & 0.14 & 0.58 \\ 
			      $\gamma_{4}$ & -0.07 & 0.07 & 0.20 & 0.79 \\ 
 			     $\gamma_{5}$ & -0.09 & 0.03 & 0.14 & 0.65 \\ 
    			     $\gamma_{6}$ & -0.02 & 0.12 & 0.26 & 0.92 \\ 
   			     $\gamma_{7}$ & -0.15 & -0.03 & 0.08 & 0.33 \\ 
  	  		  $\gamma_{8}$ & -0.14 & -0.01 & 0.12 & 0.44 \\ 
  			  $\gamma_{9}$ & -0.11 & 0.02 & 0.15 & 0.59 \\ 
 			   $\gamma_{10}$ & -0.13 & -0.01 & 0.11 & 0.43 \\ 
   			  $\gamma_{11}$ & -0.19 & -0.07 & 0.06 & 0.19 \\ 
  			   $\gamma_{12}$ & -0.13 & 0.00 & 0.13 & 0.51 \\ 
     			  $\gamma_{13}$ & -0.21 & -0.08 & 0.05 & 0.16 \\ 
			\bottomrule
		\end{tabular}}
	 \centering 
	\resizebox{0.63\textwidth}{!}{\normalsize
	\begin{tabular}{rl}
		\toprule
		$j$ & Metrics, Y\\ 
		\midrule
		1 & Education index \\ 
		2 & Employment to popn. ratio, 15+, total (\%) \\ 
		3 & GNI per capita (2011 PPP\$) \\ 
		4 & Internet users (\% of popn.) \\ 
		5 & Life expectancy at birth, total (years) \\ 
		6 & Mortality rate, infant (per 1,000 live births) \\ 
		7 & Population density \\ 
		8 & Popn. with at least some secondary education (\% ages 25 and older) \\ 
		9 & Popn., ages 65 and older (\% of total) \\ 
		10 & Popn., urban (\% of total) \\ 
		11 & Proportion of seats held by women in national parliaments (\%) \\ 
		12 & Unemployment, total (\% of total labor force) \\ 
		13 & POLITY index \\ 
		14 & Corruption Perception Index \\ 
		15 & Renewable energy consumption (\% of total final energy consumption) \\ 
		\bottomrule 
			\toprule
			$K + Q$ & Previous years' covariates, $Z^* = (X^*, Y^*)$ \\ 
			\midrule  
      1 & Access to electricity, rural (\% of rural population) \\ 
      2 & Employment to population ratio, 15+, total (\%) \\ 
      3 & Forest area (\% of land area) \\ 
      4 & Mean years of schooling (years) \\ 
      5 & Population density \\ 
      6 & Population, ages 65 and older (\% of total) \\ 
      7 & Population, total \\ 
      8 & Population, urban (\% of total) \\ 
      9 & Renewable energy consumption (\% of total final energy consumption) \\ 
      10 & Proportion of seats held by women in national parliaments (\%) \\ 
      11 & Unemployment, total (\% of total labor force) \\ 
      12 & POLITY index \\ 
      13 & CPI \\ 
			\bottomrule
			\end{tabular}}
		\end{table}

 \begin{landscape}
 	\begin{table}
    \caption{Posterior median of latent health of countries in LACSH using MML as treatment variable, and associated 90\% credible intervals}
 		\normalsize 
	\begin{tabular}[htb!]{cccc}
		\toprule
		$H_i$ & 5\% & 50\% & 95\% \\  \midrule
    $H_{1}$ & -0.11 & -0.08 & -0.05 \\ 
    $H_{2}$ & 0.03 & 0.06 & 0.08 \\ 
    $H_{3}$ & -0.15 & -0.09 & -0.04 \\ 
    $H_{4}$ & -0.01 & 0.02 & 0.04 \\ 
    $H_{5}$ & 0.06 & 0.09 & 0.13 \\ 
    $H_{6}$ & 0.06 & 0.09 & 0.12 \\ 
    $H_{7}$ & -0.04 & -0.02 & 0.01 \\ 
    $H_{8}$ & -0.12 & -0.07 & -0.04 \\ 
    $H_{9}$ & 0.05 & 0.07 & 0.10 \\ 
    $H_{10}$ & -0.11 & -0.08 & -0.05 \\ 
    $H_{11}$ & -0.14 & -0.10 & -0.07 \\ 
    $H_{12}$ & -0.07 & -0.04 & -0.01 \\ 
    $H_{13}$ & 0.05 & 0.08 & 0.11 \\ 
    $H_{14}$ & -0.15 & -0.08 & -0.03 \\ 
    $H_{15}$ & 0.04 & 0.06 & 0.10 \\ 
    $H_{16}$ & -0.00 & 0.02 & 0.04 \\ 
    $H_{17}$ & -0.08 & -0.05 & -0.02 \\ 
    $H_{18}$ & -0.08 & -0.05 & -0.02 \\ 
    $H_{19}$ & 0.06 & 0.09 & 0.13 \\ 
    $H_{20}$ & 0.06 & 0.08 & 0.11 \\ 
    $H_{21}$ & 0.03 & 0.06 & 0.09 \\ 
    $H_{22}$ & -0.01 & 0.01 & 0.03 \\ 
    $H_{23}$ & -0.10 & -0.06 & -0.04 \\ 
    $H_{24}$ & -0.10 & -0.06 & -0.03 \\ 
    $H_{25}$ & -0.02 & 0.00 & 0.02 \\ 
    $H_{26}$ & 0.02 & 0.04 & 0.06 \\ 
    $H_{27}$ & 0.02 & 0.05 & 0.07 \\ 
    $H_{28}$ & 0.06 & 0.09 & 0.13 \\ 
    $H_{29}$ & 0.07 & 0.09 & 0.13 \\ 
    $H_{30}$ & -0.02 & -0.00 & 0.02 \\ 
		\bottomrule
	\end{tabular} \hspace{.2em}
	\begin{tabular}{cccc}
		\toprule
		& 5\% & 50\% & 95\%  \\ \midrule
    $H_{31}$ & -0.04 & -0.01 & 0.02 \\ 
    $H_{32}$ & -0.02 & 0.00 & 0.02 \\ 
    $H_{33}$ & 0.05 & 0.07 & 0.10 \\ 
    $H_{34}$ & 0.07 & 0.09 & 0.13 \\ 
    $H_{35}$ & -0.11 & -0.08 & -0.05 \\ 
    $H_{36}$ & 0.07 & 0.10 & 0.14 \\ 
    $H_{37}$ & 0.05 & 0.08 & 0.11 \\ 
    $H_{38}$ & -0.07 & -0.04 & -0.01 \\ 
    $H_{39}$ & 0.05 & 0.07 & 0.11 \\ 
    $H_{40}$ & -0.09 & -0.06 & -0.03 \\ 
    $H_{41}$ & -0.12 & -0.09 & -0.06 \\ 
    $H_{42}$ & 0.04 & 0.07 & 0.10 \\ 
    $H_{43}$ & -0.03 & -0.01 & 0.01 \\ 
    $H_{44}$ & -0.04 & -0.02 & 0.01 \\ 
    $H_{45}$ & -0.04 & -0.02 & 0.01 \\ 
    $H_{46}$ & 0.05 & 0.08 & 0.11 \\ 
    $H_{47}$ & 0.06 & 0.08 & 0.12 \\ 
    $H_{48}$ & -0.04 & -0.02 & 0.00 \\ 
    $H_{49}$ & -0.05 & -0.03 & -0.00 \\ 
    $H_{50}$ & 0.04 & 0.07 & 0.10 \\ 
    $H_{51}$ & -0.07 & -0.03 & 0.00 \\ 
    $H_{52}$ & -0.09 & -0.05 & -0.02 \\ 
    $H_{53}$ & 0.01 & 0.03 & 0.06 \\ 
    $H_{54}$ & 0.06 & 0.08 & 0.12 \\ 
    $H_{55}$ & -0.01 & 0.01 & 0.03 \\ 
    $H_{56}$ & -0.05 & -0.02 & 0.01 \\ 
    $H_{57}$ & 0.07 & 0.10 & 0.14 \\ 
    $H_{58}$ & -0.02 & 0.01 & 0.04 \\ 
    $H_{59}$ & -0.09 & -0.06 & -0.04 \\ 
    $H_{60}$ & -0.06 & -0.03 & -0.01 \\ 
		\bottomrule
	\end{tabular} \hspace{.2em}
	\begin{tabular}{cccc}
		\toprule
		& 5\% & 50\% & 95\%  \\ \midrule
    $H_{61}$ & 0.03 & 0.06 & 0.09 \\ 
    $H_{62}$ & -0.13 & -0.07 & -0.03 \\ 
    $H_{63}$ & -0.08 & -0.05 & -0.02 \\ 
    $H_{64}$ & -0.01 & 0.01 & 0.04 \\ 
    $H_{65}$ & -0.10 & -0.06 & -0.02 \\ 
    $H_{66}$ & -0.01 & 0.02 & 0.05 \\ 
    $H_{67}$ & -0.13 & -0.08 & -0.05 \\ 
    $H_{68}$ & 0.05 & 0.08 & 0.11 \\ 
    $H_{69}$ & 0.05 & 0.07 & 0.11 \\ 
    $H_{70}$ & 0.05 & 0.07 & 0.11 \\ 
    $H_{71}$ & -0.04 & -0.02 & 0.01 \\ 
    $H_{72}$ & -0.00 & 0.02 & 0.05 \\ 
    $H_{73}$ & -0.15 & -0.10 & -0.07 \\ 
    $H_{74}$ & -0.05 & -0.03 & 0.00 \\ 
    $H_{75}$ & -0.03 & 0.01 & 0.04 \\ 
    $H_{76}$ & -0.13 & -0.09 & -0.05 \\ 
    $H_{77}$ & -0.02 & 0.01 & 0.03 \\ 
    $H_{78}$ & -0.12 & -0.08 & -0.05 \\ 
    $H_{79}$ & -0.02 & 0.00 & 0.02 \\ 
    $H_{80}$ & -0.10 & -0.06 & -0.04 \\ 
    $H_{81}$ & -0.13 & -0.09 & -0.05 \\ 
    $H_{82}$ & -0.02 & -0.00 & 0.02 \\ 
    $H_{83}$ & 0.05 & 0.08 & 0.11 \\ 
    $H_{84}$ & 0.07 & 0.10 & 0.14 \\ 
    $H_{85}$ & -0.06 & -0.04 & -0.01 \\ 
    $H_{86}$ & 0.06 & 0.09 & 0.12 \\ 
    $H_{87}$ & -0.14 & -0.08 & -0.03 \\ 
    $H_{88}$ & -0.09 & -0.06 & -0.03 \\ 
    $H_{89}$ & 0.00 & 0.02 & 0.04 \\ 
    $H_{90}$ & -0.00 & 0.02 & 0.05 \\ 
		\bottomrule
	\end{tabular} \hspace{0.2em}
	\begin{tabular}{cccc}
		\toprule
		& 5\% & 50\% & 95\%  \\ \midrule
    $H_{91}$ & -0.03 & -0.01 & 0.01 \\ 
    $H_{92}$ & -0.10 & -0.07 & -0.04 \\ 
    $H_{93}$ & 0.05 & 0.07 & 0.10 \\ 
    $H_{94}$ & 0.04 & 0.06 & 0.09 \\ 
    $H_{95}$ & -0.01 & 0.01 & 0.04 \\ 
    $H_{96}$ & -0.16 & -0.09 & -0.04 \\ 
    $H_{97}$ & 0.02 & 0.05 & 0.08 \\ 
    $H_{98}$ & -0.11 & -0.07 & -0.04 \\ 
    $H_{99}$ & -0.10 & -0.05 & -0.01 \\ 
    $H_{100}$ & -0.10 & -0.06 & -0.03 \\ 
    $H_{101}$ & -0.10 & -0.07 & -0.04 \\ 
    $H_{102}$ & -0.02 & -0.00 & 0.02 \\ 
    $H_{103}$ & 0.06 & 0.09 & 0.12 \\ 
    $H_{104}$ & 0.07 & 0.10 & 0.14 \\ 
    $H_{105}$ & -0.11 & -0.07 & -0.04 \\ 
    $H_{106}$ & -0.02 & 0.00 & 0.03 \\ 
    $H_{107}$ & -0.07 & -0.04 & -0.01 \\ 
    $H_{108}$ & -0.02 & 0.01 & 0.04 \\ 
    $H_{109}$ & -0.02 & 0.00 & 0.02 \\ 
    $H_{110}$ & -0.02 & 0.00 & 0.02 \\ 
    $H_{111}$ & -0.11 & -0.07 & -0.05 \\ 
    $H_{112}$ & 0.02 & 0.05 & 0.08 \\ 
    $H_{113}$ & 0.03 & 0.06 & 0.09 \\ 
    $H_{114}$ & 0.04 & 0.06 & 0.09 \\ 
    $H_{115}$ & -0.07 & -0.04 & -0.01 \\ 
    $H_{116}$ & -0.04 & -0.01 & 0.01 \\ 
    $H_{117}$ & -0.13 & -0.09 & -0.06 \\ 
    $H_{118}$ & -0.07 & -0.03 & -0.00 \\ 
    $H_{119}$ & -0.11 & -0.07 & -0.04 \\ 
    $H_{120}$ & -0.09 & -0.06 & -0.03 \\ 
		\bottomrule
	\end{tabular}
\end{table}
\end{landscape}

\newpage
\begin{landscape}
\begin{table}[htb!]
\caption{Countries used in LACSH model}\label{sec:cty}
\resizebox{1.12\textwidth}{!}{\Large
	\begin{tabular}{rl}
		\toprule
		$i$ & Country \\ 
		\midrule
		1 & Afghanistan \\ 
		2 & Albania \\ 
		3 & United Arab Emirates  \\ 
		4 & Armenia \\ 
		5 & Australia \\ 
		6 & Austria \\ 
		7 & Azerbaijan \\ 
		8 & Burundi (Anchor country) \\ 
		9 & Belgium \\ 
		10 & Benin \\ 
		11 & Burkina Faso \\ 
		12 & Bangladesh \\ 
		13 & Bulgaria \\ 
		14 & Bahrain \\ 
		15 & Belarus \\ 
		16 & Brazil \\ 
		17 & Bhutan \\ 
		18 & Botswana \\ 
		19 & Canada \\ 
		20 & Switzerland \\ 
		21 & Chile \\ 
		22 & China \\ 
		23 & Cameroon \\ 
		24 & Congo, Rep. \\ 
		25 & Colombia \\ 
		26 & Costa Rica \\ 
		27 & Cyprus \\ 
		28 & Germany \\ 
		29 & Denmark \\ 
		30 & Dominican Republic \\ 
		\bottomrule
	\end{tabular} \hspace{0.5em}
	\begin{tabular}{rl}
		\midrule
		$i$ & Country \\ 
		\midrule
		31 & Algeria \\
		32 & Ecuador \\ 
		33 & Spain \\ 
		34 & Estonia \\ 
		35 & Ethiopia \\ 
		36 & Finland \\ 
		37 & France \\ 
		38 & Gabon \\ 
		39 & United Kingdom \\ 
		40 & Ghana \\ 
		41 & Gambia, The \\ 
		42 & Greece \\ 
		43 & Guatemala \\ 
		44 & Guyana \\ 
		45 & Honduras \\ 
		46 & Croatia \\ 
		47 & Hungary \\ 
		48 & Indonesia \\ 
		49 & India \\ 
		50 & Ireland \\ 
		51 & Iran, Islamic Rep. \\ 
		52 & Iraq \\ 
		53 & Israel \\ 
		54 & Italy \\ 
		55 & Jamaica \\ 
		56 & Jordan \\ 
		57 & Japan \\ 
		58 & Kazakhstan \\ 
		59 & Kenya \\ 
		60 & Cambodia \\
		\bottomrule
	\end{tabular} \hspace{0.5em}
	\begin{tabular}{rl}
		\toprule
		$i$ & Country \\ 
		\midrule
		61 & South Korea \\ 
		62 & Kuwait \\ 
		63 & Lao PDR \\ 
		64 & Lebanon \\ 
		65 & Liberia \\ 
		66 & Sri Lanka \\ 
		67 & Lesotho \\ 
		68 & Lithuania \\ 
		69 & Luxembourg \\ 
		70 & Latvia \\ 
		71 & Morocco \\ 
		72 & Mexico \\ 
		73 & Mali \\ 
		74 & Myanmar \\ 
		75 & Mongolia \\ 
		76 & Mozambique \\ 
		77 & Mauritius \\ 
		78 & Malawi \\ 
		79 & Malaysia \\ 
		80 & Namibia \\ 
		81 & Niger \\ 
		82 & Nicaragua \\ 
		83 & Netherlands \\ 
		84 & Norway \\ 
		85 & Nepal \\ 
		86 & New Zealand \\ 
		87 & Oman \\ 
		88 & Pakistan \\ 
		89 & Panama \\ 
		90 & Peru \\
		\bottomrule
	\end{tabular} \hspace{0.5em}
	\begin{tabular}{rl}
		\toprule
		$i$ & Country \\ 
		\midrule
		91 & Philippines \\ 
		92 & Papua New Guinea \\ 
		93 & Poland \\ 
		94 & Portugal \\ 
		95 & Paraguay \\ 
		96 & Qatar \\ 
		97 & Russian Federation \\ 
		98 & Rwanda \\ 
		99 & Saudi Arabia \\ 
		100 & Sudan \\ 
		101 & Senegal \\ 
		102 & El Salvador \\ 
		103 & Slovenia \\ 
		104 & Sweden \\ 
		105 & Togo \\ 
		106 & Thailand \\ 
		107 & Tajikistan \\ 
		108 & Trinidad and Tobago \\ 
		109 & Tunisia \\ 
		110 & Turkey \\ 
		111 & Uganda \\ 
		112 & Ukraine \\ 
		113 & Uruguay \\ 
		114 & United States \\ 
		115 & Uzbekistan \\ 
		116 & Vietnam \\ 
		117 & Yemen, Rep. \\ 
		118 & South Africa \\ 
		119 & Zambia \\ 
		120 & Zimbabwe \\ 
		\bottomrule
	\end{tabular}}
\end{table}
\vfill
\end{landscape}

\begin{landscape}
    \begin{table}[htb!]
    \centering
    \caption{Posterior median ranks of countries in LACSH using MML as the treatment variable, and associated 90\% credible intervals}\label{sec:cty2}
    \resizebox{1.15\textwidth}{!}{\large
      \begin{tabular}{rlll}
        \toprule
Country & 5\% & 50\% & 95\% \\ 
\midrule
Finland & 1 & 4 & 14 \\ 
Norway & 1 & 4 & 16 \\ 
Sweden & 1 & 4 & 15 \\ 
Japan & 1 & 5 & 20 \\ 
Canada & 1 & 8 & 24 \\ 
Denmark & 2 & 8 & 20 \\ 
Estonia & 2 & 9 & 22 \\ 
Australia & 1 & 10 & 28 \\ 
Austria & 3 & 10 & 22 \\ 
Germany & 2 & 10 & 22 \\ 
New Zealand & 3 & 11 & 27 \\ 
Slovenia & 3 & 12 & 25 \\ 
Italy & 2 & 14 & 29 \\ 
Hungary & 4 & 15 & 28 \\ 
Switzerland & 6 & 17 & 30 \\ 
Lithuania & 7 & 18 & 31 \\ 
Netherlands & 6 & 19 & 32 \\ 
Bulgaria & 5 & 20 & 32 \\ 
France & 8 & 20 & 31 \\ 
Croatia & 7 & 20 & 32 \\ 
Luxembourg & 7 & 20 & 32 \\ 
Latvia & 7 & 20 & 33 \\ 
United Kingdom & 9 & 22 & 33 \\ 
Spain & 7 & 23 & 35 \\ 
Poland & 10 & 23 & 33 \\ 
Belgium & 11 & 24 & 34 \\ 
Ireland & 12 & 26 & 36 \\ 
Greece & 11 & 28 & 37 \\ 
Belarus & 12 & 29 & 38 \\ 
Portugal & 14 & 29 & 38 \\ 
United States & 17 & 30 & 38 \\ 
South Korea & 15 & 31 & 39 \\ 
Chile & 19 & 32 & 40 \\ 
Albania & 20 & 33 & 39 \\ 
Uruguay & 15 & 33 & 42 \\ 
Russian Federation & 18 & 35 & 43 \\ 
Cyprus & 27 & 36 & 43 \\ 
Ukraine & 24 & 37 & 44 \\ 
Costa Rica & 33 & 40 & 48 \\ 
Israel & 33 & 41 & 53 \\ 
            \bottomrule
    \end{tabular} \hspace{0.5em}
    \begin{tabular}{rlll}
    	\midrule
    	Country & 5\% & 50\% & 95\% \\ 
    \midrule
 Peru & 37 & 44 & 58 \\ 
 Mexico & 37 & 45 & 57 \\ 
Panama & 39 & 45 & 55 \\ 
Armenia & 39 & 47 & 62 \\ 
Brazil & 39 & 47 & 62 \\ 
Sri Lanka & 34 & 47 & 68 \\ 
Lebanon & 39 & 49 & 67 \\ 
Paraguay & 39 & 49 & 65 \\ 
Trinidad and Tobago & 37 & 50 & 70 \\ 
Kazakhstan & 39 & 52 & 72 \\ 
Mauritius & 41 & 52 & 69 \\ 
China & 42 & 53 & 67 \\ 
Jamaica & 44 & 54 & 68 \\ 
Mongolia & 40 & 54 & 75 \\ 
Ecuador & 45 & 56 & 69 \\ 
Tunisia & 44 & 56 & 70 \\ 
Colombia & 46 & 57 & 70 \\ 
Malaysia & 45 & 57 & 73 \\ 
Thailand & 43 & 57 & 72 \\ 
Turkey & 46 & 58 & 71 \\ 
Nicaragua & 46 & 60 & 74 \\ 
El Salvador & 48 & 60 & 73 \\ 
Dominican Republic & 48 & 61 & 74 \\ 
Philippines & 50 & 65 & 77 \\ 
Algeria & 50 & 67 & 84 \\ 
Guatemala & 52 & 67 & 79 \\ 
Vietnam & 53 & 68 & 80 \\ 
Azerbaijan & 52 & 69 & 85 \\ 
Honduras & 54 & 69 & 81 \\ 
Guyana & 52 & 70 & 83 \\ 
Jordan & 54 & 70 & 86 \\ 
Indonesia & 57 & 71 & 82 \\ 
Morocco & 56 & 71 & 84 \\ 
India & 59 & 75 & 87 \\ 
Myanmar & 57 & 75 & 89 \\ 
Iran, Islamic Rep. & 59 & 76 & 100 \\ 
Cambodia & 63 & 78 & 92 \\ 
South Africa & 60 & 79 & 98 \\ 
Nepal & 65 & 80 & 94 \\ 
Uzbekistan & 64 & 80 & 101 \\ 
    \bottomrule
    \end{tabular} \hspace{0.5em}
    \begin{tabular}{rlll}
    	\midrule
    	Country & 5\% & 50\% & 95\% \\ 
    \midrule
Gabon & 63 & 81 & 97 \\ 
Bangladesh & 69 & 82 & 96 \\ 
Tajikistan & 69 & 82 & 97 \\ 
Bhutan & 71 & 86 & 101 \\ 
Botswana & 71 & 87 & 106 \\ 
Iraq & 74 & 88 & 109 \\ 
Lao PDR & 74 & 88 & 102 \\ 
Saudi Arabia & 69 & 90 & 114 \\ 
Pakistan & 79 & 92 & 107 \\ 
Ghana & 79 & 93 & 107 \\ 
Liberia & 73 & 93 & 111 \\ 
Zimbabwe & 78 & 93 & 107 \\ 
Kenya & 81 & 94 & 108 \\ 
Congo, Rep. & 81 & 95 & 110 \\ 
Sudan & 81 & 96 & 112 \\ 
Cameroon & 82 & 98 & 112 \\ 
Namibia & 81 & 98 & 114 \\ 
Papua New Guinea & 83 & 99 & 114 \\ 
Senegal & 84 & 99 & 111 \\ 
Kuwait & 77 & 101 & 117 \\ 
Zambia & 88 & 102 & 114 \\ 
Togo & 88 & 103 & 114 \\ 
Uganda & 88 & 104 & 115 \\ 
Burundi & 82 & 105 & 118 \\ 
Rwanda & 86 & 105 & 117 \\ 
Benin & 90 & 106 & 116 \\ 
Ethiopia & 90 & 106 & 116 \\ 
Oman & 80 & 107 & 119 \\ 
Afghanistan & 94 & 108 & 117 \\ 
Bahrain & 77 & 108 & 119 \\ 
Malawi & 93 & 108 & 117 \\ 
Gambia, The & 98 & 111 & 118 \\ 
Lesotho & 90 & 111 & 120 \\ 
Niger & 91 & 111 & 119 \\ 
Yemen, Rep. & 93 & 112 & 119 \\ 
United Arab Emirates & 82 & 113 & 120 \\ 
Mozambique & 95 & 113 & 120 \\ 
Qatar & 84 & 115 & 120 \\ 
Burkina Faso & 106 & 117 & 120 \\ 
Mali & 107 & 117 & 120 \\ 
    \bottomrule
       \end{tabular}}
    \end{table}
     \end{landscape}
     
\subsection{Posterior summaries for LACSH model, GGHE as treatment variable}\label{ssec:appC2}\phantomsection{\hspace{1em}}
	\vspace{1em}
	\begin{table}[hbt!]
   \caption{Posterior median of parameters in LACSH using GGHE as the treatment variable, and associated 90\% credible intervals} \
		\label{tab:paramB2}
		\centering
	\resizebox{0.9\textwidth}{!}{\normalsize
		\begin{tabular}{rrrr}
			\toprule
			$a_j$\tablefootnote{Refer to Appendix \ref{ssec:appC1} for indexing of metrics and covariates} & 5\% & 50\% & 95\% \\
			\midrule
 		$a_{1}$ & 14.48 & 16.81 & 19.23 \\ 
$a_{2}$ & -6.53 & -3.70 & -0.90 \\ 
$a_{3}$ & 15.43 & 17.81 & 20.23 \\ 
$a_{4}$ & 14.97 & 17.30 & 19.73 \\ 
$a_{5}$ & 14.64 & 16.97 & 19.45 \\ 
$a_{6}$ & 15.10 & 17.41 & 19.83 \\ 
$a_{7}$ & -2.48 & 0.39 & 3.28 \\ 
$a_{8}$ & 11.92 & 14.32 & 16.87 \\ 
$a_{9}$ & 11.38 & 13.86 & 16.53 \\ 
$a_{10}$ & 11.82 & 14.26 & 16.88 \\ 
$a_{11}$ & -0.96 & 1.88 & 4.76 \\ 
$a_{12}$ & -3.18 & -0.34 & 2.56 \\ 
$a_{13}$ & 4.87 & 7.60 & 10.49 \\ 
$a_{14}$ & 11.67 & 14.18 & 16.90 \\ 
$a_{15}$ & -13.64 & -10.86 & -8.26 \\ 
			\midrule
			$\sigma_H^2$ & 0.002 & 0.003 & 0.004 \\ 
			$\phi$ & 27.10 & 29.64 & 33.16  \\ 
			\midrule
			$\Sigma_{Y\{2,12\}}$ \tablefootnote{Only the top five in magnitude of the posterior median are presented.} & 0.71 & 0.55 & 0.71  \\
			$\Sigma_{Y\{9,13\}}$ & 0.36 & 0.26 & 0.36 \\
			$\Sigma_{Y\{3,15\}}$ &  0.36 & 0.25 & 0.36\\ 
			$\Sigma_{Y\{7,12\}}$ & 0.29 & 0.15 & 0.29 \\
			$\Sigma_{Y\{11,13\}}$ & 0.28 & 0.15 & 0.28\\ 
			\bottomrule
		\end{tabular}
		\hspace{1.5em}
		\begin{tabular}{rrrrc}
			\toprule 
			$\beta_k$ & 5\% & 50\% & 95\% &  p($\beta > 0 \ |$ data) \\
			\midrule
			$\beta_{0}$ & -0.07 & 0.02 & 0.13 & 0.63 \\ 
			$\beta_{1}$ & 0.04 & 0.04 & 0.05 & 1.00 \\ 
			$\beta_{2}$ & 0.00 & 0.00 & 0.01 & 0.99 \\ 
			$\beta_{3}$ & -0.01 & 0.01 & 0.03 & 0.75 \\ 
			$\beta_{4}$ & -0.03 & -0.01 & 0.02 & 0.35 \\ 
			$\beta_{5}$ & 0.00 & 0.01 & 0.02 & 0.98 \\
			\midrule
			$\gamma_{(1+K+Q)}$ & 5\% & 50\% & 95\% &  p($\gamma > 0 \ |$  data) \\
			\midrule 
   			$\gamma_{0}$ & -0.14 & -0.00 & 0.14 & 0.50 \\ 
			$\gamma_{1}$ & -0.04 & 0.11 & 0.26 & 0.90 \\ 
			$\gamma_{2}$ & -0.16 & -0.01 & 0.13 & 0.44 \\ 
			$\gamma_{3}$ & -0.14 & 0.01 & 0.15 & 0.53 \\ 
			$\gamma_{4}$ & -0.02 & 0.13 & 0.28 & 0.92 \\ 
			$\gamma_{5}$ & -0.15 & -0.01 & 0.13 & 0.43 \\ 
			$\gamma_{6}$ & -0.06 & 0.09 & 0.24 & 0.83 \\ 
			$\gamma_{7}$ & -0.16 & -0.03 & 0.11 & 0.38 \\ 
			$\gamma_{8}$ & -0.03 & 0.12 & 0.27 & 0.91 \\ 
			$\gamma_{9}$ & -0.24 & -0.09 & 0.06 & 0.15 \\ 
			$\gamma_{10}$ & -0.13 & 0.01 & 0.15 & 0.56 \\ 
			$\gamma_{11}$ & -0.15 & -0.01 & 0.13 & 0.46 \\ 
			$\gamma_{12}$ & -0.09 & 0.05 & 0.20 & 0.73 \\ 
			$\gamma_{13}$ & -0.03 & 0.12 & 0.27 & 0.91 \\ 
			\bottomrule
		\end{tabular}} \vspace{0.5em}
		\centering 
	\end{table}
\clearpage

 \begin{landscape}
	\begin{table}
   \caption{Posterior median of latent health of countries in LACSH using GGHE as the treatment variable, and associated 90\% credible intervals}
	\normalsize
	\begin{tabular}[htb!]{cccc}
		\toprule
		$H_i$ & 5\% & 50\% & 95\% \\  \midrule
	$H_{1}$ & -0.10 & -0.08 & -0.07 \\ 
$H_{2}$ & -0.00 & 0.00 & 0.01 \\ 
$H_{3}$ & 0.03 & 0.04 & 0.06 \\ 
$H_{4}$ & -0.02 & -0.01 & 0.00 \\ 
$H_{5}$ & 0.06 & 0.08 & 0.09 \\ 
$H_{6}$ & 0.06 & 0.07 & 0.08 \\ 
$H_{7}$ & -0.01 & -0.00 & 0.01 \\ 
$H_{8}$ & -0.11 & -0.09 & -0.07 \\ 
$H_{9}$ & 0.06 & 0.07 & 0.08 \\ 
$H_{10}$ & -0.09 & -0.08 & -0.07 \\ 
$H_{11}$ & -0.10 & -0.09 & -0.07 \\ 
$H_{12}$ & -0.07 & -0.06 & -0.04 \\ 
$H_{13}$ & 0.02 & 0.03 & 0.04 \\ 
$H_{14}$ & 0.02 & 0.04 & 0.05 \\ 
$H_{15}$ & 0.03 & 0.04 & 0.05 \\ 
$H_{16}$ & -0.00 & 0.01 & 0.02 \\ 
$H_{17}$ & -0.02 & -0.01 & -0.00 \\ 
$H_{18}$ & -0.03 & -0.02 & -0.01 \\ 
$H_{19}$ & 0.06 & 0.07 & 0.08 \\ 
$H_{20}$ & 0.06 & 0.08 & 0.09 \\ 
$H_{21}$ & 0.03 & 0.04 & 0.05 \\ 
$H_{22}$ & 0.00 & 0.01 & 0.02 \\ 
$H_{23}$ & -0.08 & -0.07 & -0.06 \\ 
$H_{24}$ & -0.07 & -0.06 & -0.04 \\ 
$H_{25}$ & -0.00 & 0.01 & 0.02 \\ 
$H_{26}$ & 0.01 & 0.02 & 0.03 \\ 
$H_{27}$ & 0.03 & 0.04 & 0.05 \\ 
$H_{28}$ & 0.06 & 0.08 & 0.09 \\ 
$H_{29}$ & 0.06 & 0.08 & 0.09 \\ 
$H_{30}$ & -0.01 & -0.00 & 0.00 \\
		\bottomrule
	\end{tabular}\hspace{0.2em}
	\begin{tabular}{cccc}
		\toprule
		& 5\% & 50\% & 95\%  \\ \midrule
$H_{31}$ & 0.00 & 0.01 & 0.02 \\ 
$H_{32}$ & -0.01 & 0.00 & 0.01 \\ 
$H_{33}$ & 0.05 & 0.06 & 0.07 \\ 
$H_{34}$ & 0.04 & 0.05 & 0.07 \\ 
$H_{35}$ & -0.10 & -0.08 & -0.07 \\ 
$H_{36}$ & 0.07 & 0.08 & 0.09 \\ 
$H_{37}$ & 0.06 & 0.07 & 0.08 \\ 
$H_{38}$ & -0.03 & -0.02 & -0.01 \\ 
$H_{39}$ & 0.06 & 0.07 & 0.08 \\ 
$H_{40}$ & -0.07 & -0.05 & -0.04 \\ 
$H_{41}$ & -0.10 & -0.09 & -0.07 \\ 
$H_{42}$ & 0.03 & 0.04 & 0.06 \\ 
$H_{43}$ & -0.04 & -0.03 & -0.02 \\ 
$H_{44}$ & -0.04 & -0.03 & -0.02 \\ 
$H_{45}$ & -0.05 & -0.04 & -0.03 \\ 
$H_{46}$ & 0.03 & 0.04 & 0.05 \\ 
$H_{47}$ & 0.03 & 0.04 & 0.05 \\ 
$H_{48}$ & -0.03 & -0.02 & -0.01 \\ 
$H_{49}$ & -0.05 & -0.04 & -0.03 \\ 
$H_{50}$ & 0.06 & 0.07 & 0.09 \\ 
$H_{51}$ & 0.01 & 0.02 & 0.03 \\ 
$H_{52}$ & -0.04 & -0.03 & -0.01 \\ 
$H_{53}$ & 0.04 & 0.05 & 0.07 \\ 
$H_{54}$ & 0.04 & 0.05 & 0.07 \\ 
$H_{55}$ & -0.02 & -0.01 & -0.00 \\ 
$H_{56}$ & -0.01 & -0.00 & 0.01 \\ 
$H_{57}$ & 0.07 & 0.08 & 0.09 \\ 
$H_{58}$ & 0.00 & 0.02 & 0.03 \\ 
$H_{59}$ & -0.08 & -0.07 & -0.05 \\ 
$H_{60}$ & -0.06 & -0.04 & -0.03 \\    
		\bottomrule
	\end{tabular}\hspace{0.2em}
	\begin{tabular}{cccc}
		\toprule
		& 5\% & 50\% & 95\%  \\ \midrule
$H_{61}$ & 0.05 & 0.06 & 0.07 \\ 
$H_{62}$ & 0.03 & 0.05 & 0.06 \\ 
$H_{63}$ & -0.05 & -0.04 & -0.03 \\ 
$H_{64}$ & 0.01 & 0.02 & 0.03 \\ 
$H_{65}$ & -0.11 & -0.09 & -0.08 \\ 
$H_{66}$ & -0.01 & 0.00 & 0.01 \\ 
$H_{67}$ & -0.09 & -0.07 & -0.06 \\ 
$H_{68}$ & 0.04 & 0.05 & 0.06 \\ 
$H_{69}$ & 0.07 & 0.08 & 0.09 \\ 
$H_{70}$ & 0.03 & 0.04 & 0.05 \\ 
$H_{71}$ & -0.02 & -0.01 & -0.00 \\ 
$H_{72}$ & 0.00 & 0.01 & 0.02 \\ 
$H_{73}$ & -0.11 & -0.09 & -0.08 \\ 
$H_{74}$ & -0.05 & -0.04 & -0.03 \\ 
$H_{75}$ & -0.02 & -0.01 & 0.01 \\ 
$H_{76}$ & -0.13 & -0.11 & -0.09 \\ 
$H_{77}$ & 0.00 & 0.01 & 0.02 \\ 
$H_{78}$ & -0.11 & -0.09 & -0.08 \\ 
$H_{79}$ & 0.02 & 0.03 & 0.04 \\ 
$H_{80}$ & -0.04 & -0.03 & -0.02 \\ 
$H_{81}$ & -0.11 & -0.09 & -0.08 \\ 
$H_{82}$ & -0.03 & -0.02 & -0.01 \\ 
$H_{83}$ & 0.06 & 0.07 & 0.09 \\ 
$H_{84}$ & 0.07 & 0.08 & 0.10 \\ 
$H_{85}$ & -0.07 & -0.06 & -0.05 \\ 
$H_{86}$ & 0.06 & 0.07 & 0.08 \\ 
$H_{87}$ & 0.02 & 0.03 & 0.05 \\ 
$H_{88}$ & -0.07 & -0.06 & -0.05 \\ 
$H_{89}$ & 0.01 & 0.02 & 0.03 \\ 
$H_{90}$ & -0.01 & 0.00 & 0.01 \\ 
		\bottomrule
	\end{tabular} \hspace{0.2em}
	\begin{tabular}{cccc}
		\toprule
		& 5\% & 50\% & 95\%  \\ \midrule
   $H_{91}$ & -0.03 & -0.02 & -0.01 \\ 
$H_{92}$ & -0.07 & -0.05 & -0.04 \\ 
$H_{93}$ & 0.04 & 0.05 & 0.06 \\ 
$H_{94}$ & 0.04 & 0.05 & 0.06 \\ 
$H_{95}$ & -0.02 & -0.01 & 0.00 \\ 
$H_{96}$ & 0.04 & 0.06 & 0.08 \\ 
$H_{97}$ & 0.02 & 0.04 & 0.05 \\ 
$H_{98}$ & -0.08 & -0.07 & -0.06 \\ 
$H_{99}$ & 0.03 & 0.04 & 0.06 \\ 
$H_{100}$ & -0.08 & -0.06 & -0.05 \\ 
$H_{101}$ & -0.08 & -0.06 & -0.05 \\ 
$H_{102}$ & -0.02 & -0.01 & -0.00 \\ 
$H_{103}$ & 0.04 & 0.05 & 0.06 \\ 
$H_{104}$ & 0.07 & 0.08 & 0.10 \\ 
$H_{105}$ & -0.10 & -0.08 & -0.07 \\ 
$H_{106}$ & 0.01 & 0.02 & 0.03 \\ 
$H_{107}$ & -0.07 & -0.06 & -0.05 \\ 
$H_{108}$ & 0.01 & 0.02 & 0.03 \\ 
$H_{109}$ & -0.01 & 0.00 & 0.01 \\ 
$H_{110}$ & 0.02 & 0.03 & 0.04 \\ 
$H_{111}$ & -0.10 & -0.08 & -0.07 \\ 
$H_{112}$ & -0.01 & 0.00 & 0.01 \\ 
$H_{113}$ & 0.03 & 0.04 & 0.05 \\ 
$H_{114}$ & 0.06 & 0.08 & 0.09 \\ 
$H_{115}$ & -0.04 & -0.03 & -0.02 \\ 
$H_{116}$ & -0.03 & -0.02 & -0.01 \\ 
$H_{117}$ & -0.09 & -0.08 & -0.06 \\ 
$H_{118}$ & -0.04 & -0.02 & -0.01 \\ 
$H_{119}$ & -0.08 & -0.07 & -0.05 \\ 
$H_{120}$ & -0.09 & -0.08 & -0.06 \\
     \bottomrule
	\end{tabular}
	\end{table}
\end{landscape}

\begin{landscape}
    \begin{table}[htb!]
    \centering
    \caption{Posterior median ranks of countries in LACSH using GGHE as the treatment variable, and associated 90\% credible intervals}\label{sec:cty3}
    \resizebox{1.15\textwidth}{!}{\large
      \begin{tabular}{rlll}
        \toprule
Country & 5\% & 50\% & 95\% \\ 
\midrule
Norway & 1 & 2 & 10 \\ 
Sweden & 1 & 4 & 11 \\ 
Australia & 1 & 5 & 16 \\ 
Japan & 1 & 5 & 15 \\ 
Finland & 2 & 7 & 15 \\ 
Luxembourg & 1 & 7 & 14 \\ 
Switzerland & 2 & 8 & 16 \\ 
Germany & 2 & 8 & 15 \\ 
Denmark & 2 & 8 & 16 \\ 
United States & 2 & 9 & 17 \\ 
Netherlands & 4 & 11 & 17 \\ 
Ireland & 4 & 12 & 18 \\ 
Canada & 4 & 13 & 20 \\ 
Austria & 6 & 14 & 19 \\ 
Belgium & 6 & 14 & 18 \\ 
United Kingdom & 6 & 14 & 19 \\ 
New Zealand & 5 & 15 & 21 \\ 
France & 9 & 16 & 20 \\ 
South Korea & 16 & 21 & 29 \\ 
Qatar & 10 & 21 & 33 \\ 
Spain & 17 & 22 & 29 \\ 
Estonia & 19 & 23 & 31 \\ 
Israel & 18 & 23 & 29 \\ 
Italy & 18 & 23 & 31 \\ 
Portugal & 19 & 25 & 33 \\ 
Slovenia & 20 & 25 & 32 \\ 
Lithuania & 20 & 26 & 33 \\ 
Poland & 23 & 29 & 37 \\ 
Greece & 24 & 31 & 39 \\ 
Kuwait & 21 & 31 & 41 \\ 
Saudi Arabia & 22 & 32 & 41 \\ 
Hungary & 26 & 33 & 40 \\ 
United Arab Emirates & 23 & 34 & 44 \\ 
Cyprus & 27 & 34 & 41 \\ 
Chile & 26 & 35 & 43 \\ 
Croatia & 29 & 36 & 42 \\ 
Latvia & 28 & 36 & 42 \\ 
Uruguay & 27 & 37 & 45 \\ 
Bahrain & 26 & 38 & 46 \\ 
Belarus & 30 & 38 & 45 \\ 
            \bottomrule
    \end{tabular} \hspace{0.5em}
    \begin{tabular}{rlll}
    	\midrule
    	Country & 5\% & 50\% & 95\% \\ 
    \midrule
Russian Federation & 30 & 39 & 46 \\ 
Oman & 32 & 42 & 51 \\ 
Bulgaria & 36 & 43 & 48 \\ 
Malaysia & 37 & 43 & 49 \\ 
Turkey & 37 & 43 & 48 \\ 
Costa Rica & 41 & 47 & 54 \\ 
Lebanon & 43 & 48 & 54 \\ 
Trinidad and Tobago & 41 & 49 & 58 \\ 
Panama & 44 & 50 & 57 \\ 
Thailand & 45 & 51 & 59 \\ 
Iran, Islamic Rep. & 45 & 52 & 60 \\ 
Kazakhstan & 45 & 52 & 60 \\ 
Mauritius & 46 & 53 & 62 \\ 
Algeria & 47 & 54 & 62 \\ 
Mexico & 46 & 54 & 62 \\ 
China & 48 & 55 & 62 \\ 
Brazil & 49 & 56 & 65 \\ 
Colombia & 52 & 59 & 66 \\ 
Albania & 53 & 60 & 67 \\ 
Peru & 52 & 61 & 69 \\ 
Ecuador & 55 & 62 & 69 \\ 
Tunisia & 54 & 62 & 69 \\ 
Ukraine & 53 & 62 & 69 \\ 
Sri Lanka & 53 & 64 & 73 \\ 
Azerbaijan & 57 & 65 & 73 \\ 
Jordan & 57 & 65 & 72 \\ 
Dominican Republic & 60 & 67 & 74 \\ 
Mongolia & 59 & 69 & 77 \\ 
Paraguay & 61 & 69 & 76 \\ 
Armenia & 63 & 70 & 77 \\ 
El Salvador & 64 & 70 & 76 \\ 
Jamaica & 65 & 71 & 78 \\ 
Bhutan & 64 & 72 & 80 \\ 
Morocco & 66 & 73 & 80 \\ 
Gabon & 69 & 76 & 84 \\ 
Indonesia & 70 & 77 & 83 \\ 
Vietnam & 71 & 77 & 84 \\ 
Philippines & 71 & 78 & 85 \\ 
Botswana & 73 & 80 & 87 \\ 
Nicaragua & 74 & 80 & 85 \\ 
   \bottomrule
    \end{tabular} \hspace{0.5em}
    \begin{tabular}{rlll}
    	\midrule
    	Country & 5\% & 50\% & 95\% \\ 
    \midrule
South Africa & 73 & 80 & 87 \\ 
Iraq & 75 & 82 & 88 \\ 
Namibia & 75 & 83 & 89 \\ 
Uzbekistan & 75 & 83 & 88 \\ 
Guatemala & 78 & 84 & 88 \\ 
Guyana & 77 & 84 & 89 \\ 
Honduras & 82 & 87 & 90 \\ 
India & 84 & 88 & 92 \\ 
Myanmar & 84 & 89 & 92 \\ 
Lao PDR & 86 & 90 & 94 \\ 
Cambodia & 87 & 91 & 94 \\ 
Ghana & 90 & 94 & 100 \\ 
Papua New Guinea & 89 & 94 & 100 \\ 
Bangladesh & 91 & 95 & 101 \\ 
Congo, Rep. & 91 & 96 & 102 \\ 
Nepal & 92 & 96 & 102 \\ 
Tajikistan & 91 & 96 & 103 \\ 
Pakistan & 92 & 97 & 103 \\ 
Sudan & 93 & 99 & 105 \\ 
Senegal & 94 & 100 & 106 \\ 
Kenya & 95 & 101 & 106 \\ 
Zambia & 96 & 101 & 106 \\ 
Rwanda & 96 & 103 & 108 \\ 
Cameroon & 98 & 104 & 109 \\ 
Lesotho & 97 & 105 & 112 \\ 
Zimbabwe & 102 & 106 & 112 \\ 
Yemen, Rep. & 101 & 107 & 116 \\ 
Benin & 104 & 109 & 115 \\ 
Uganda & 105 & 110 & 117 \\ 
Togo & 106 & 111 & 116 \\ 
Afghanistan & 106 & 112 & 118 \\ 
Ethiopia & 106 & 112 & 118 \\ 
Gambia, The & 107 & 113 & 118 \\ 
Burundi & 105 & 114 & 119 \\ 
Burkina Faso & 107 & 114 & 118 \\ 
Malawi & 109 & 115 & 119 \\ 
Liberia & 109 & 116 & 119 \\ 
Mali & 110 & 116 & 119 \\ 
Niger & 113 & 118 & 119 \\ 
Mozambique & 119 & 120 & 120 \\ 
     \bottomrule
       \end{tabular}}
    \end{table}
 \end{landscape}

\section{Ladder plot of rankings between LACSH and selected existing methods}\label{sec:ladder1} \vspace{-0.9em}
The left of the ladder plot in Figure \ref{fig:MMLHDI} shows the country ranking according to the posterior median of $H$ from \textit{LACSH} with MML as treatment variable. The right of the plot is the HDI ranking from the year 2016, and `NIS' denotes countries that are not in our data for LACSH. Generally, the two sets of rankings are somewhat similar, except for some of the high-income countries such as Saudi Arabia, Kuwait, and Oman that have drastically different ranks between the indices. Thus, MML may be viewed as a driver of human development.  \vspace{-.1em}

Figure \ref{fig:MMLSPI} replaces the HDI ranking with the SPI ranking in 2016. The SPI, too, ranks a few high-income countries drastically lower than LACSH. The overall similarity between LACSH and SPI rankings suggests that MML can be seen as a driver of social progress.  \vspace{-.1em}

Figure \ref{fig:GGHEHDI} replaces the $H$'s of Figure \ref{fig:MMLHDI} with LACSH posterior median values of $H$ that utilise GGHE as treatment variable. Here, countries are ranked much more similarly between LACSH and HDI. It also shows a stronger similarity according to income group. This illustrates that not only is GGHE highly correlated with the income of a country, but it also can be seen as a driver of human development.  \vspace{-.1em}

Finally, Figure \ref{fig:GGHESPI} displays the LACSH rankings with the SPI ranking in 2016. Again, we see that the countries are ranked quite similarly, except for a few high-income countries that are ranked lower in the SPI ranking. 

\begin{figure}[H]
	\begin{center}
	\caption{Ladder plot for (posterior median of) $H$'s between using MML and using GGHE as the treatment variable}
	\includegraphics[width=1.2\textwidth, height=\textheight]{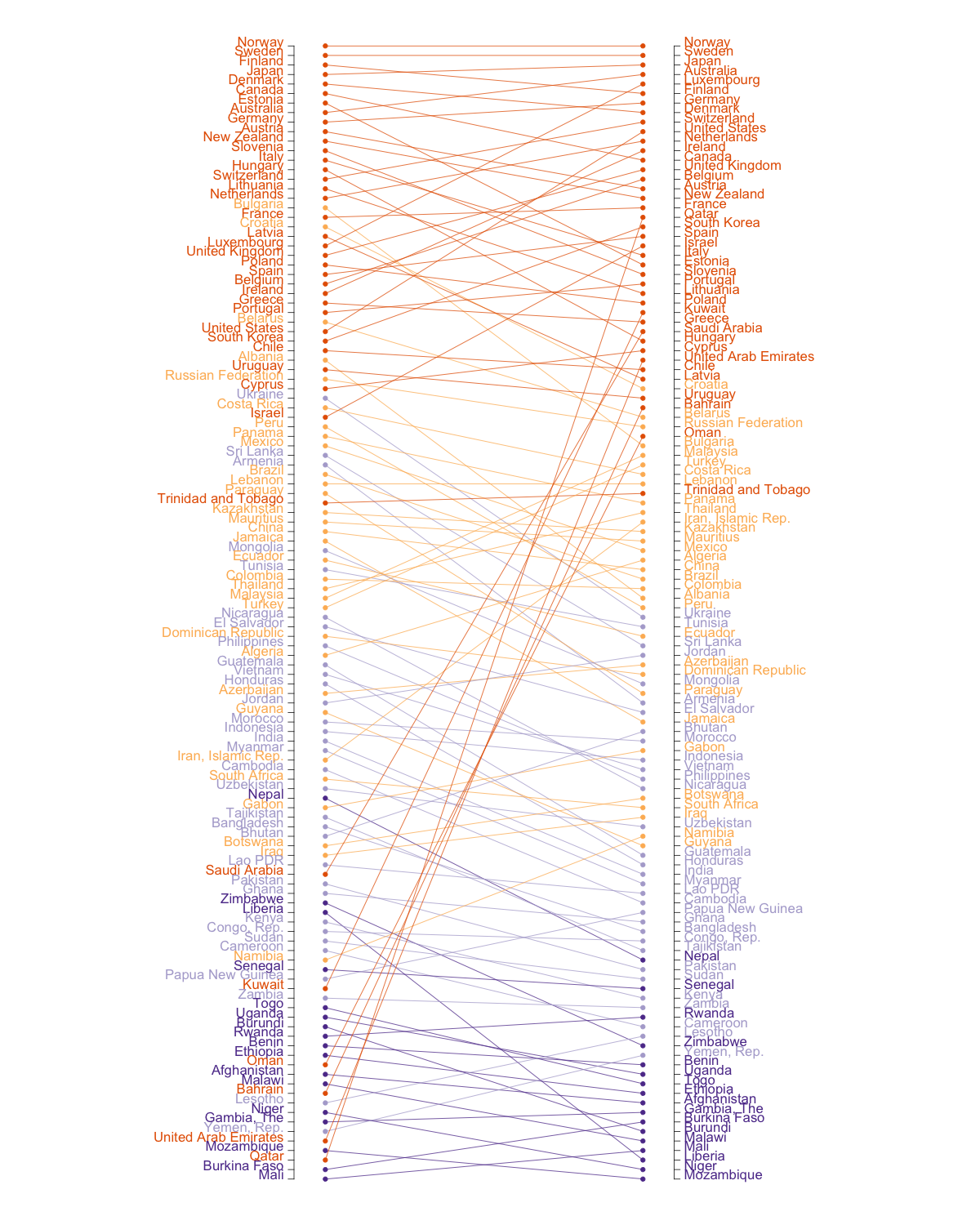}  
	\label{fig:MML2.1GGHE}
  \end{center}
\end{figure}\vspace{-2em}

\begin{figure}[H]
	\centering
	\caption{Ladder plot for (posterior median of) latent $H$ ranking from LACSH using MML as the treatment variable against HDI ranking }
	\includegraphics[width=0.9\textwidth, height=\textheight]{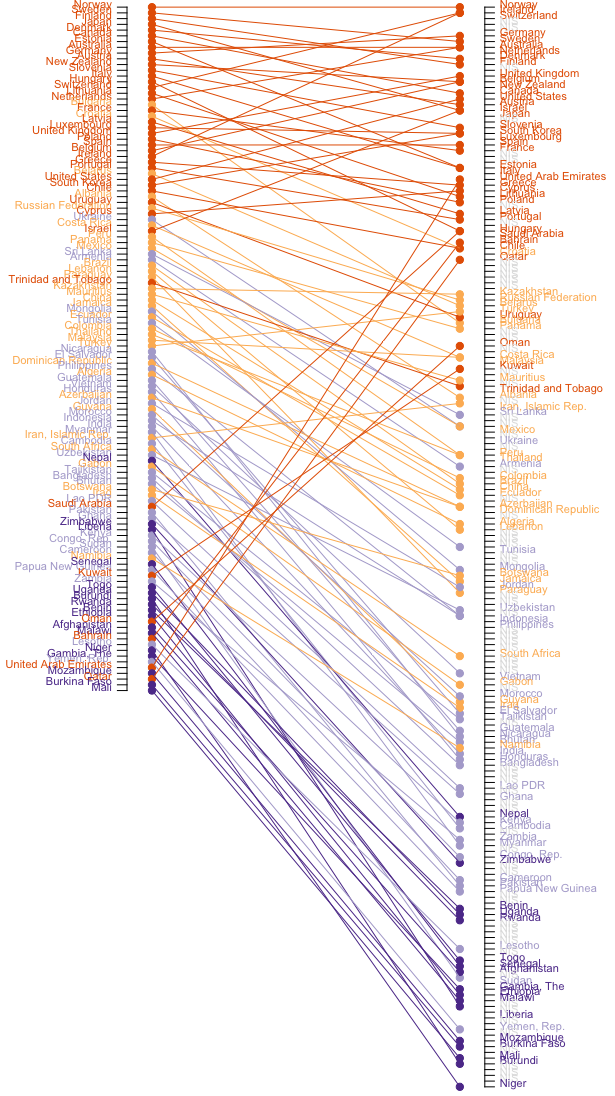}  \label{fig:MMLHDI}
\end{figure}\vspace{-2em}

\begin{figure}[H]
	\centering
	\caption{Ladder plot for (posterior median of) latent $H$ ranking from LACSH using MML as the treatment variable against SPI ranking } 
	\includegraphics[width=0.9\textwidth, height=\textheight]{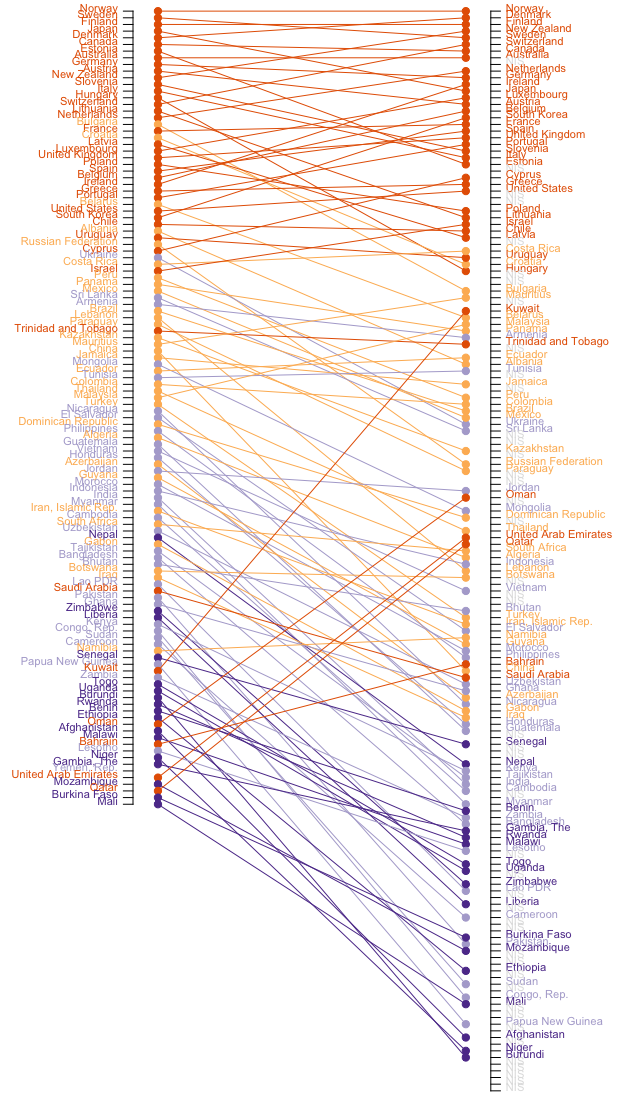} \label{fig:MMLSPI}
\end{figure}\vspace{-2em}

\begin{figure}[H]
	\centering
	\caption{Ladder plot for (posterior median of) latent $H$ ranking from LACSH using GGHE as the treatment variable against HDI ranking }
	\includegraphics[width=0.9\textwidth, height=\textheight]{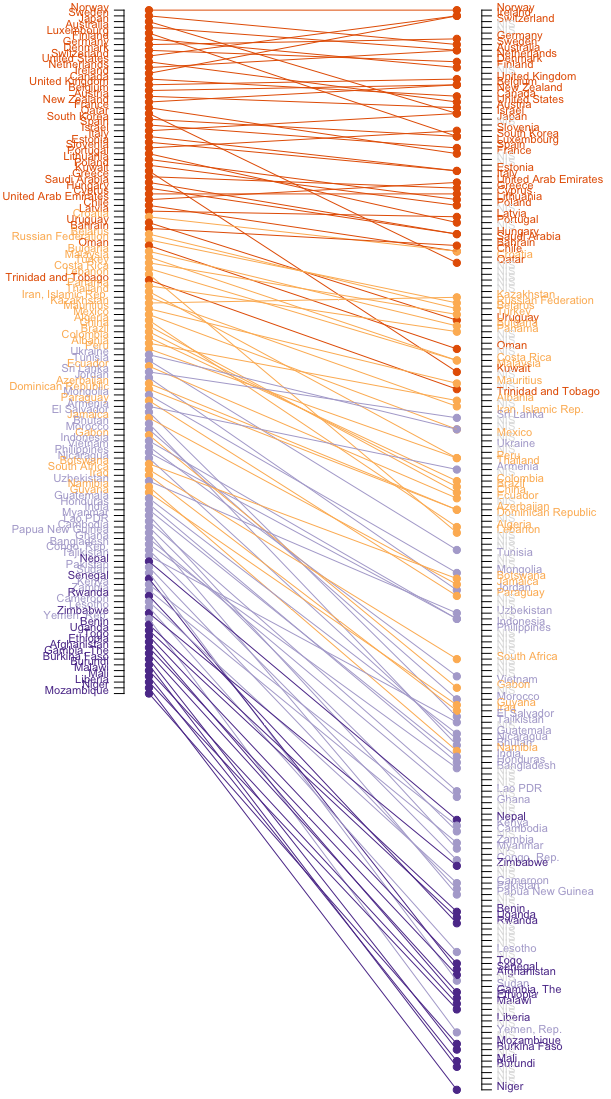} \label{fig:GGHEHDI}
\end{figure}\vspace{-2em}

\begin{figure}[H]
	\centering
	\caption{Ladder plot for (posterior median of) latent $H$ ranking from LACSH using GGHE as the treatment variable against SPI ranking }
	\includegraphics[width=0.9\textwidth, height=\textheight]{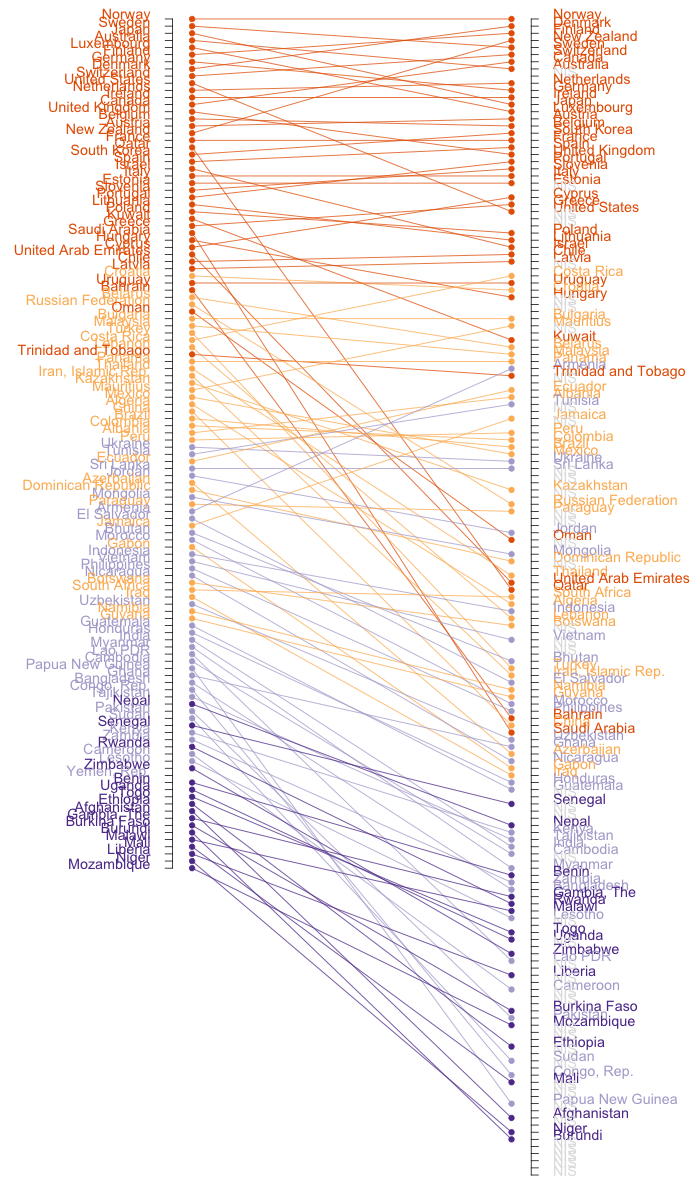} \label{fig:GGHESPI}
\end{figure}\vspace{-2em}

\section{Diagnostics: Effective sample sizes}
For model results, we utilised roughly 100,000 post-burn-in MCMC samples from the posterior distribution. Standard diagnostics suggested that each parameter of the MCMC chain had reached its steady state. We present trace plots for selected parameters from our LACSH models in the results in Table 5.1 in the main paper.

Moreover, to check convergence of our models, we calculate the effective sample size for each of the parameters. The effective sample size adjusts the total sample size due to autocorrelation, resulting in an approximate sample size that reflects what would have been an independent sample. Table \ref{tab:table1} shows the minimum and maximum effective sample size of all of the parameters calculated using the \texttt{effectiveSize} function in the \texttt{coda} package in R. 

\begin{table}[h!]
  \begin{center}
    \caption{Minimum and maximum effective size of parameters for \textit{LACSH} models} \vspace{0.5em}
    \label{tab:table1}
    \begin{tabular}{c|c|c} 
    \toprule
     \multirow{2}{*}{\textbf{Models}} & \multirow{2}{*}{\parbox{4cm}{\textbf{Minimum effective sample size}}} & \multirow{2}{*}{\parbox{4cm}{\textbf{Maximum effective sample size}}} \\ \\
      \midrule
       {MML as treatment} & 256.04 & 126927.2\\
       {GGHE as treatment} & 542.06 & 385777.8\\
       \bottomrule
    \end{tabular}
  \end{center}
\end{table}

\section{Additional interesting results on parameters}\label{appendix:appC} \vspace{-1em}
\begin{figure}[H]
	\includegraphics[width=0.825\textwidth, height=0.3\textheight]{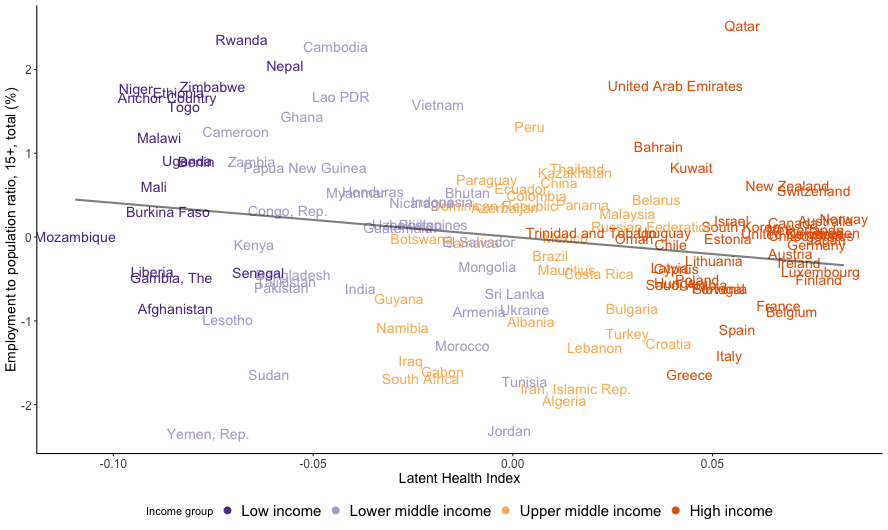}
	\vspace{-0.8em}
		\caption{Plot of employment-to-population ratio  vs. posterior median of latent health, with GGHE as treatment and with a least squares regression fit for visualization}
	\label{fig:employ}
\end{figure}
Figure \ref{fig:employ} shows a negative relationship for employment-to-population ratio and countries' latent health, conditioned on other metrics with GGHE as treatment (90\% credible interval for $a_{2}$ is $(-6.53, -0.90)$). The relevant discussions appear in Section 5.3.

\begin{figure}[H]
	\includegraphics[width=0.825\textwidth,height=0.3\textheight]{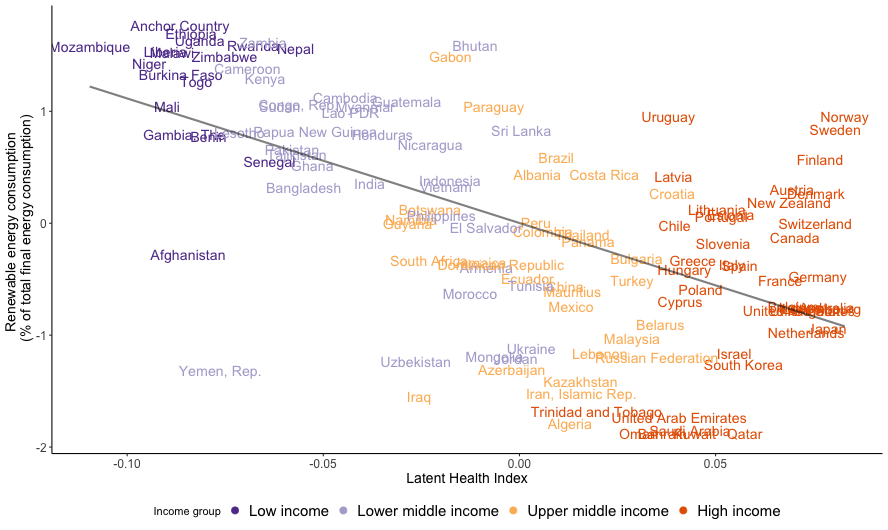}\vspace{-0.8em}
	\caption{Plot of renewable energy consumption vs. posterior median of latent health, with GGHE as treatment and with a least squares regression fit for visualization}
	\label{fig:renew}
\end{figure}

Additionally, with GGHE as treatment, the metric `\texttt{renewable energy consumption}' (shown in Figure \ref{fig:renew} and discussed in Section 6.3) also has a negative relationship with the country's latent health (90\% credible interval for $a_{15}$ is $(-13.64, -8.26)$).

\vspace{-1.5em}
There are a few possible explanations for this negative relationship. For example, lower-income countries lack the capital to expand existing infrastructure for electricity access, especially into rural areas, hence certain renewable energy sources that do not rely on existing infrastructure serve as more viable options. In addition, high-income countries may be reluctant, for various reasons, to transition from established infrastructure to renewable energy sources.
\vspace{-1em}
\section{Posterior average dose-response curves for subsample of countries}\label{appendix:appE} 
\vspace{-1em}
\begin{figure}[!ht]
	\includegraphics[width=0.8\textwidth,height=0.28\textheight]{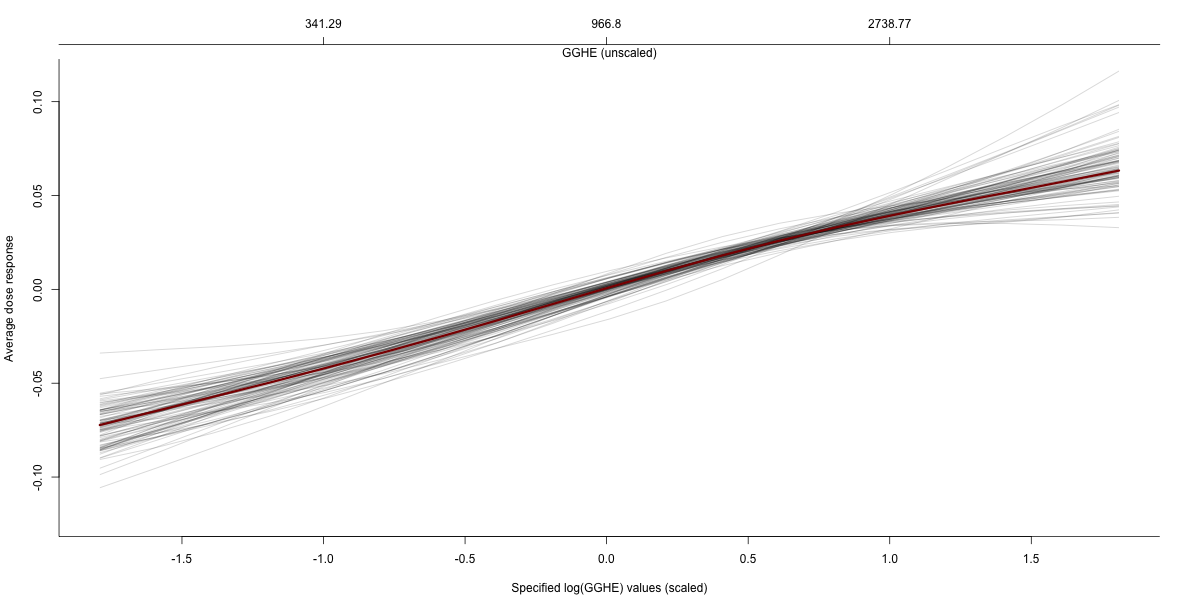} \vspace{-1em}
	\caption{Posterior median dose-response function (red), with GGHE as treatment, for the subsample of 82 countries, with 100 thinned MCMC samples (gray) for visualization of uncertainty}
	\label{fig:subDR}
\end{figure}

Figure \ref{fig:subDR} shows the average dose-response curves for 82 countries, a subsample of the original set of 120 countries. Subsampling was done to address the consistent pattern of covariate imbalance in Figure 10. Compared to using the full sample, the average dose-response curve for the subsample shows less uncertainty with tighter credible bands for the dose-response curve, but the MCMC samples are no longer always monotone.  

\begin{figure}[H]
	\includegraphics[width=0.8\textwidth,height=0.3\textheight]{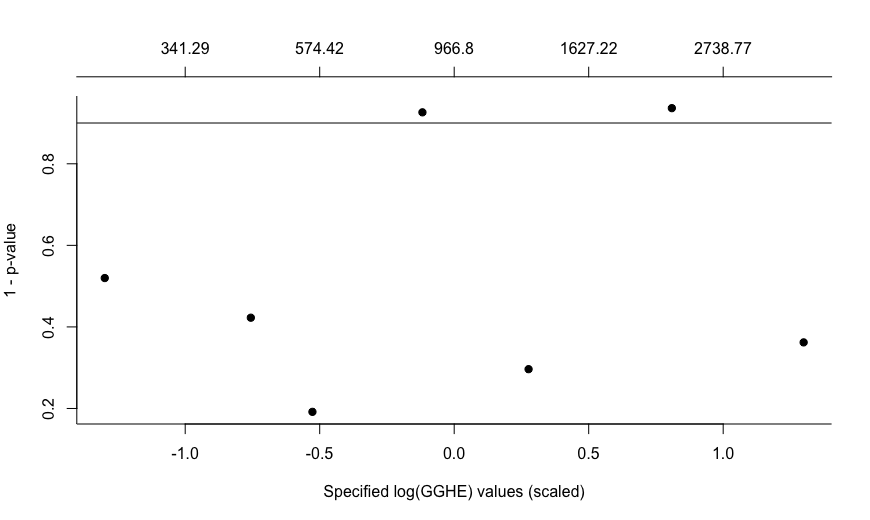}
		\caption{Plot of p-values (reversed-scale) for assessing covariate balance, with GGHE as treatment, for the subsample of 82 countries}
	\label{fig:CB_DR}
\end{figure}
Figure \ref{fig:CB_DR} no longer shows any consistent pattern of covariate imbalance.

\newpage
\section{Simulation studies}\label{appendix:simStudies} 

\subsection{Simulation study I: Cross validation study}
\begin{figure}[H]
	\includegraphics[height=0.5\textheight]{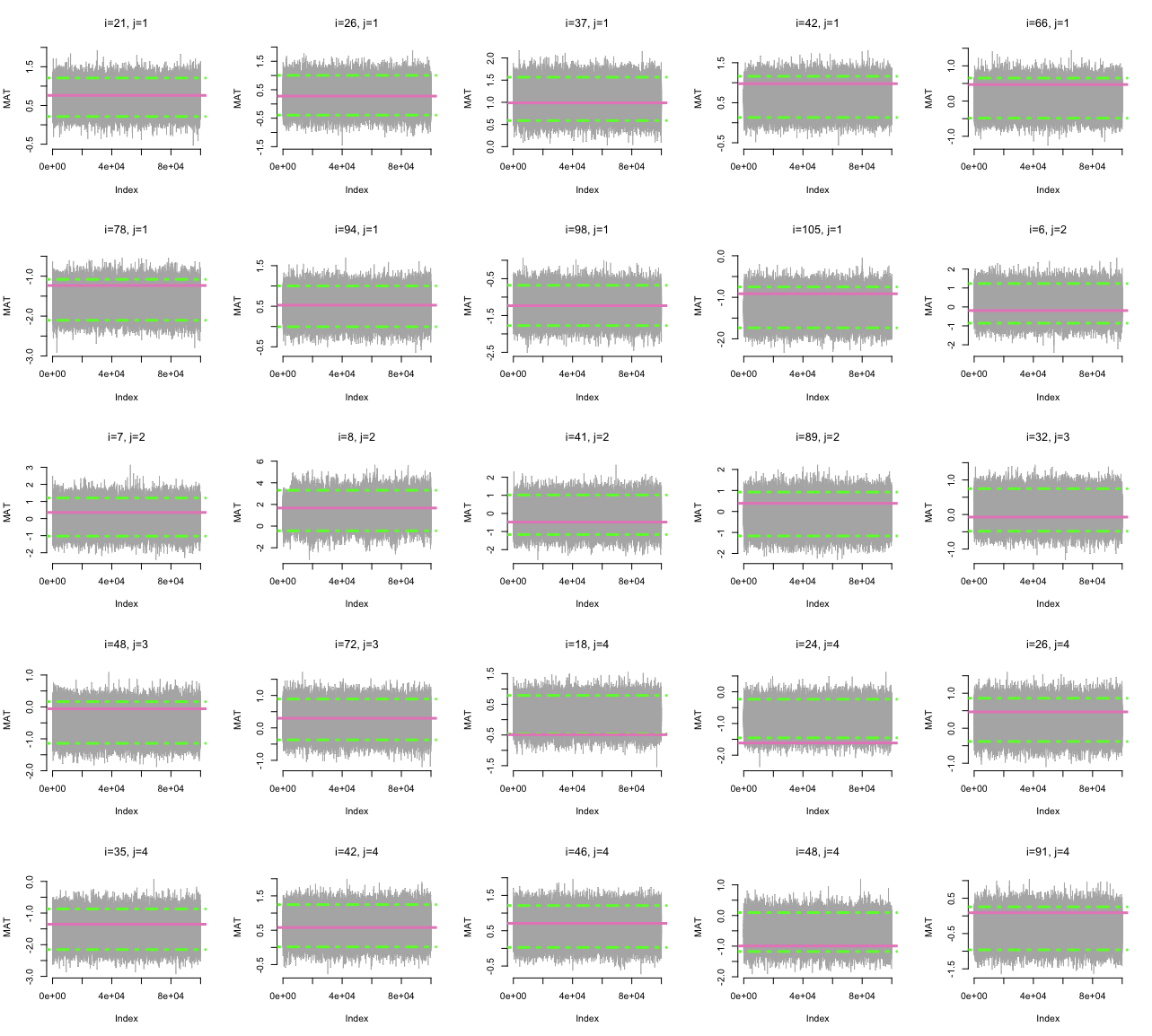}
		\caption{Trace plots of the corresponding missing $y_{ij}$'s for a chosen iteration to be representative of all 10 iterations.  The green dashed lines are 95\% posterior predictive intervals and the red lines represent the true $y_{ij}$'s before withholding them as missing. We find that across the 10 iterations, the lowest capture rate for missing $y_{ij}$'s values is 90\% and the highest capture rate is 98\%.}
	\label{fig:ymis}
\end{figure}

\begin{table}[!ht] \centering 
  \caption{`Capture' rate for 100 missing $y_{ij}$ values using a 95\% posterior predictive interval for each of the 10 iterations. For each iteration, we determine how many times (out of 100) the true value was `captured' by the intervals, and tabulate the rates of capture in this table.} 
  \label{} 

\begin{tabular}{@{\extracolsep{3pt}} ccccccccccc} 
\\[-1.8ex]\hline 
\hline \\[-1.8ex] 
Iteration & $1$ & $2$ & $3$ & $4$ & $5$ & $6$ & $7$ & $8$ & $9$ & $10$ \\ 
Capture rate & $0.90$ & $0.98$ & $0.91$ & $0.92$ & $0.93$ & $0.90$ & $0.98$ & $0.92$ & $0.90$ & $0.92$ \\ 
\hline \\[-1.8ex] 
\end{tabular} 
\end{table} 

\newpage
\subsection{Simulation study II: Recovering parameter values from fictitious data}

\begin{figure}[H]
	\includegraphics[height=0.35\textheight, width=0.45\textwidth]{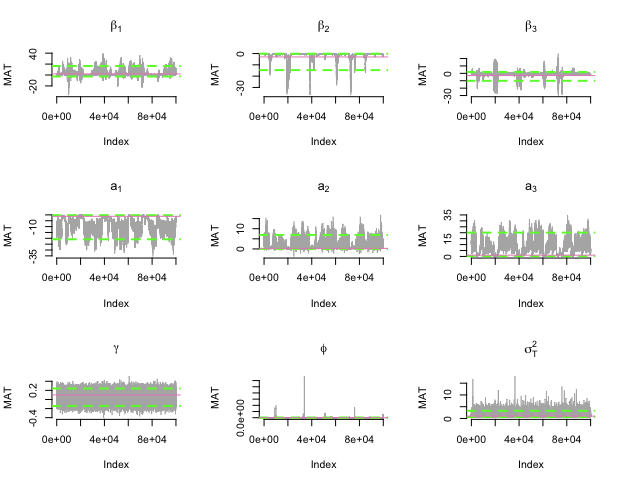}
  	\includegraphics[height=0.35\textheight,width=0.45\textwidth]{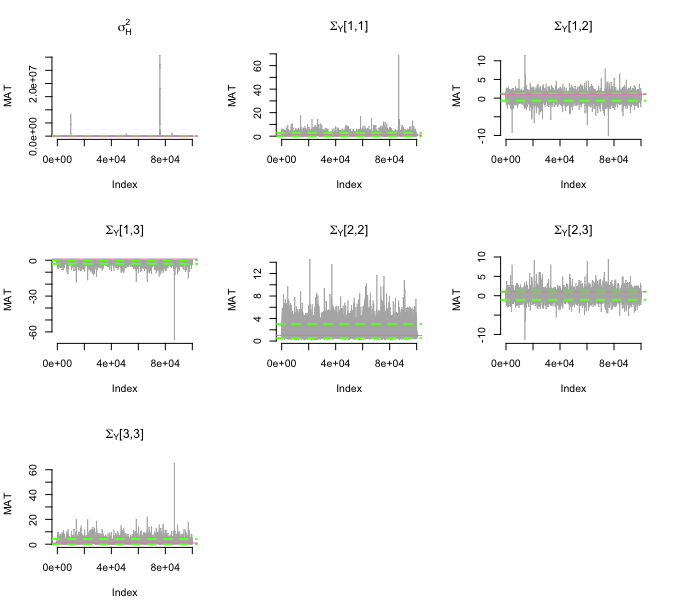} \vspace{-1em}
		\caption{Trace plots of the corresponding parameters for a chosen iteration (iteration 21 out of 50) for $\bm{\beta} = \{2, -3, -3 \}$. The green dashed lines are 95\% credible intervals and the red lines represent the true parameter values.}
	\label{fig:ygen21_1}
\end{figure}

\begin{figure}[H]
	\includegraphics[height=0.35\textheight,width=0.45\textwidth]{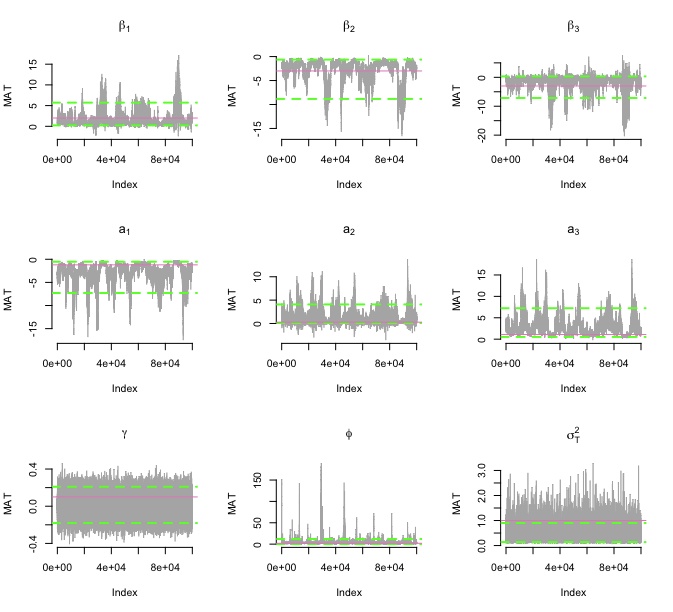}
  	\includegraphics[height=0.35\textheight,width=0.45\textwidth]{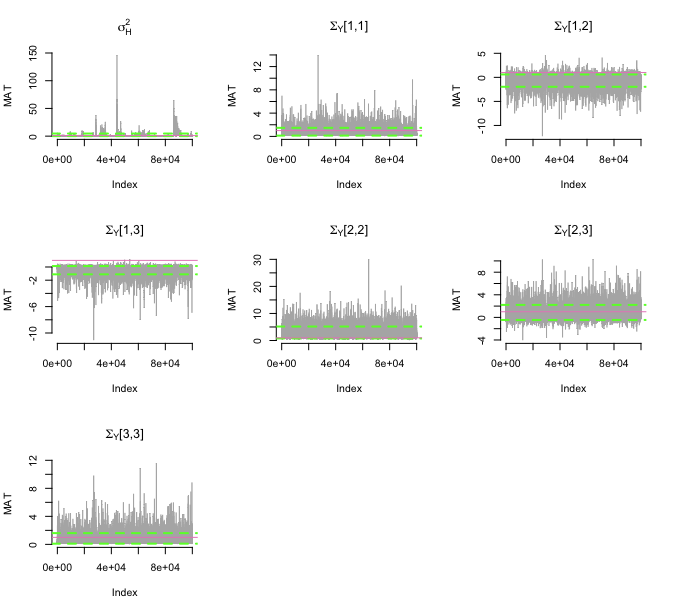} \vspace{-1em}
		\caption{Trace plots of the corresponding parameters for a chosen iteration (iteration 24 out of 50) for $\bm{\beta} = \{2, -3, -3 \}$. The green dashed lines are 95\% credible intervals and the red lines represent the true parameter values.}
	\label{fig:ygen24_1}
\end{figure}

\begin{table}[!htbp] \centering 
  \caption{`Capture' rate for each parameter using a 95\% credible interval over 50 iterations. For each iteration, the LACSH model was fitted to the randomly generated $H$'s and $Y$'s. For each model parameter, we determine how many times the true value was `captured' by the fifty 95\% credible intervals, and tabulate the rates of capture in this table.}
  \label{} 
\begin{tabular}{@{\extracolsep{5pt}} ccccc} 
\\[-1.8ex]\hline 
\hline \\[-1.8ex] 
 & Parameter & $\bm{\beta} = \{2, 3, 3\}$ & $\bm{\beta} = \{2, -3, \frac{(-\beta_0 + \beta_1 \bar{T})}{R}\}$ & $\bm{\beta} = \{2, -3, -3\}$ \\ 
\hline \\[-1.8ex] 
1 & $a_1$ & $0.84$ & $0.70$ & $0.82$ \\ 
2 & $a_2$ & $0.86$ & $0.74$ & $0.86$ \\ 
3 & $a_3$ & $0.86$ & $0.72$ & $0.80$ \\ 
4 & $\beta_1$ & $0.90$ & $1$ & $0.92$ \\ 
5 & $\beta_2$ & $0.80$ & $0.68$ & $0.82$ \\ 
6 & $\beta_3$ & $0.94$ & $0.72$ & $0.72$ \\ 
7 & $\gamma$ & $1$ & $1$ & $1$ \\ 
8 & $\phi$ & $0.86$ & $0.96$ & $0.90$ \\ 
9 & $\sigma^2_H$ & $0.90$ & $0.84$ & $0.86$ \\ 
10 & $\sigma^2_T$ & $0.94$ & $0.90$ & $0.92$ \\ 
11 & $\Sigma_{y[1,1]}$ & $0.90$ & $0.96$ & $0.92$ \\ 
12 & $\Sigma_{y[1,2]}$ & $0.98$ & $0.94$ & $0.96$ \\ 
13 & $\Sigma_{y[1,3]}$ & $0.88$ & $0.88$ & $0.90$ \\ 
14 & $\Sigma_{y[2,2]}$ & $0.94$ & $0.90$ & $0.96$ \\ 
15 & $\Sigma_{y[2,3]}$ & $1$ & $0.98$ & $0.94$ \\ 
16 & $\Sigma_{y[3,3]}$& $0.86$ & $0.80$ & $0.82$ \\ 
\hline \\[-1.8ex] 
\end{tabular} 
\end{table} 

\bibliographystyle{ba}
\bibliography{LACSH.bib}

\end{document}